\documentclass[aps,pre,longbibliography,twocolumn,floatfix,superscriptaddress,nofootinbib]{revtex4-2}  
\pdfoutput=1
\usepackage[utf8]{inputenc}
\usepackage{amssymb,graphicx,color}
\usepackage[intlimits]{amsmath}
\usepackage[english]{babel}
\usepackage[colorlinks]{hyperref}
\usepackage{tikz}
\usepackage{comment}
\usepackage{float}
\usepackage{tabularx}
\usepackage{array}
\usepackage{xcolor}
\usepackage{ragged2e}
\usepackage{natbib}

\newcolumntype{A}{>{\centering\arraybackslash \columncolor{white!50!white}}m{2.1cm}}
\newcolumntype{B}{>{\centering\arraybackslash \columncolor{white}}m{7.9cm}}
\newcolumntype{C}{>{\centering\arraybackslash \columncolor{white!50}}m{7.9cm}}
\newcolumntype{D}{>{\centering\arraybackslash \columncolor{white!42}}m}
\newcolumntype{P}[1]{>{\centering\arraybackslash}p{#1}}
\def\beq{\begin{equation}}
\def\eeq{\end{equation}}
\def\bea{\begin{eqnarray}}
\def\eea{\end{eqnarray}}
\def\barr{\begin{array}}
	\def\earr{\end{array}}

\newcommand\Tstrut{\rule{0pt}{2.6ex}}         
\definecolor{linkcolour}{rgb}{0,0,1}
\hypersetup{citecolor=blue,linkcolor=linkcolour}
\newcommand{\flc}[1]{\textcolor{linkcolour}{(#1)}}
\definecolor{crimson}{rgb}{0.86,0.08,0.24}

\begin{document}

\title{Study of the de Almeida-Thouless (AT) line in the one-dimensional diluted power-law XY spin glass}
	
\author{Bharadwaj Vedula}
\affiliation{Department of Physics, Indian Institute of Science Education and Research, Bhopal, Madhya Pradesh 462066, India}
\author{M. A. Moore}
\affiliation{Department of Physics and Astronomy, University of Manchester, Manchester M13 9PL, United Kingdom}
\author{Auditya Sharma}
\affiliation{Department of Physics, Indian Institute of Science Education and Research, Bhopal, Madhya Pradesh 462066, India}
\date{\today}
\begin{abstract}
We study the AT line in the one-dimensional power-law diluted XY spin
glass model, in which the probability that two spins separated by a
distance $r$ interact with each other, decays as $1/r^{2\sigma}$.
Tuning the exponent $\sigma$ is equivalent to changing the space
dimension of a short-range model. We develop a heat bath algorithm to
equilibrate XY spins; using this in conjunction with the standard
parallel tempering and overrelaxation sweeps, we carry out large scale
Monte Carlo simulations.  For $\sigma=0.6$, which is in the mean-field
regime above six dimensions -- it is similar to being in 10 dimensions
-- we find clear evidence for an AT line. For $\sigma=0.75$ and
$\sigma = 0.85$, which are in the non-mean-field regime and similar to
four and three dimensions respectively, our data is like that found in
previous studies of the Ising and Heisenberg spin glasses when
reducing the temperature at fixed field. For $\sigma = 0.75$, there is
evidence from finite size scaling studies for an AT transition but for
$\sigma = 0.85$, the evidence for a transition is non-existent.  We
have also studied these systems at fixed temperature varying the field
and discovered that at both $\sigma = 0.75$ and at $\sigma =0.85$
there is evidence of an AT transition! Confusingly, the correlation
length and spin glass susceptibility as a function of the field are
both entirely consistent with the predictions of the droplet picture
and hence the non-existence of an AT line. In the usual finite size
critical point scaling studies used to provide evidence for an AT
transition, there is seemingly good evidence for an AT line at $\sigma
= 0.75$ for small values of the system size $N$, which is
strengthening as $N$ is increased, but for $N > 2048$ the trend
changes and the evidence then weakens as $N$ is further increased. We
have also studied with fewer bond realizations the system at $\sigma =
0.70$, which is the analogue of a system with short-range interactions
just below six dimensions, and found that it is similar in its
behavior to the system at $\sigma = 0.75$ but with larger finite size
corrections. The evidence from our simulations points to the complete
absence of the AT line in dimensions outside the mean-field region and
to the correctness of the droplet picture. Previous simulations which
suggested there was an AT line can be attributed to the consequences
of studying systems which are just too small. The collapse of our data
to the droplet scaling form is poor for $\sigma = 0.75$ and to some
extent also for $\sigma = 0.85$, when the correlation length becomes
of the order of the length of the system, due to the existence of
excitations which only cost a free energy of $O(1)$, just as envisaged
in the TNT picture of the ordered state of spin glasses. However, for
the case of $\sigma = 0.85$ we can provide evidence that for larger
system sizes, droplet scaling will prevail even when the correlation
length is comparable to the system size.

\end{abstract}

\maketitle 

\section{Introduction}
\begin{figure}
  
  \includegraphics[width=0.48\textwidth]{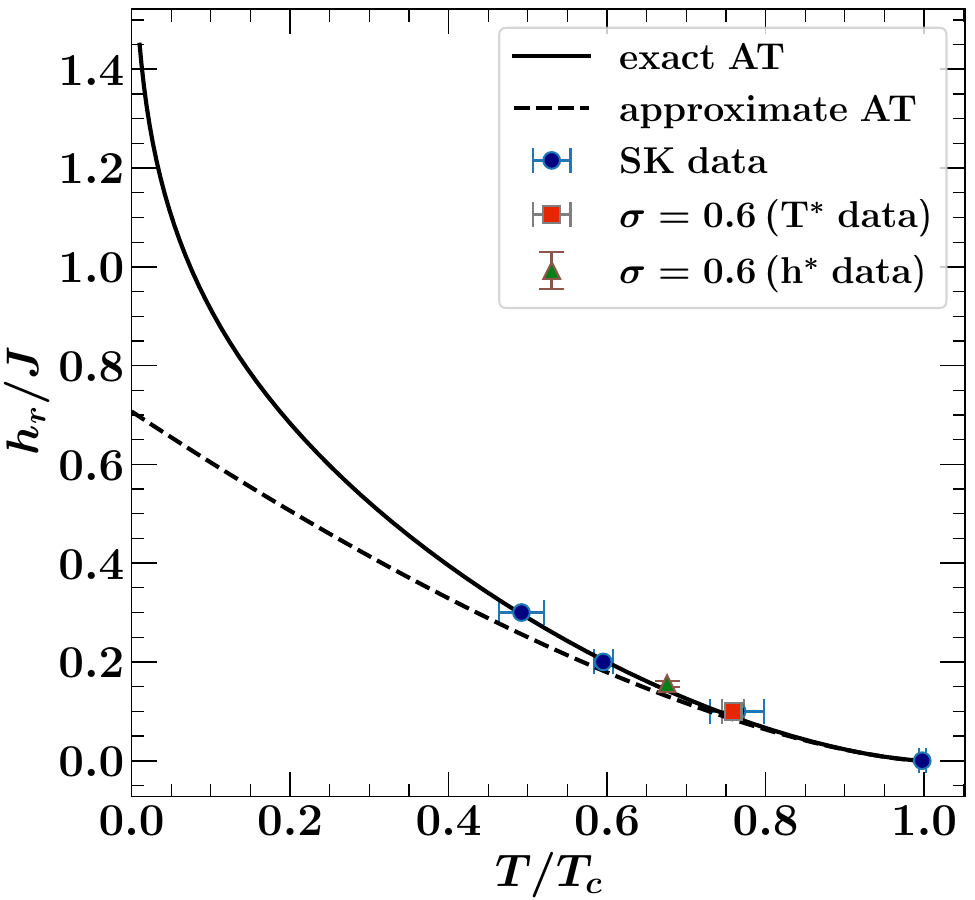}
  \caption{The AT line. The solid line is the exact AT line for the SK
    model, calculated as in Ref. \cite{sharma2010almeida}. The dashed
    line is the approximation to it of
    Eq.~(\ref{eq:approx_AT_line}). We have marked on the diagram the
    results of our simulations on the SK model, which were done to
    check our Monte Carlo procedures. The points in red and green are
    the results of our simulations at $\sigma = 0.6$, which lie in the
    mean-field region. The data on the horizontal axis for
    $\sigma = 0.6$ are normalized to the transition temperature $T_c$
    in zero field for that value of $\sigma$. For the XY SK model the
    AT line goes to infinity as $T \to 0$.}
  \label{fig:XY_AT}
\end{figure}

While the spin glass problem at mean-field level is now
well-understood~\cite{mezard1987spin}, questions remain as to the
nature of the ordered state in three dimensional spin glasses. A key
question is whether the ordered phase of real spin glasses has the
broken replica symmetry features found in mean-field theory. This
question is most easily answered by finding whether on application of
a magnetic field $h_r$ there is a line, the so-called de Almeida
Thouless (AT) line~\cite{de1978stability}, below which in the $h_r -T$
plane there is replica symmetry breaking. This line exists at
mean-field level (see Fig.~\ref{fig:XY_AT}) and its possible existence
in three dimensions can be studied experimentally and with
simulations. Simulational studies of the existence of replica symmetry
breaking within the zero-field spin glass state itself are plagued by
finite size effects: it is expected that the difference between the
predictions of droplet scaling and those of replica symmetry breaking
will only become visible for very large systems (for a review see
\cite{Moore:21}). A recent review of simulations, including studies of
the existence of the AT line, can be found in
Ref. \cite{martinmayor2022numerical}.

Right from the early days of spin glass studies there have been doubts
raised as to whether the AT line existed below six dimensions. For
example Bray and Roberts ~\cite{brayroberts:80} attempted to do an
expansion in $6-\epsilon$ dimensions for the critical exponents at the
AT line but failed to find a stable fixed point. They suggested that
maybe that indicated that there might be no AT line below six
dimensions. A renormalization group calculation also gave indications
that the AT line was going away as $d \to 6$ from above
~\cite{Moore:11}. As it is difficult to do simulations in dimensions
around $6$ to check these speculations, simulators have had to turn
instead to one-dimensional models with long-range power-law
interactions.

These models  go back to
Kotliar, Anderson and Stein \cite{PhysRevB.27.602}, who in turn were
inspired by the long-range ferromagnet that was studied by Dyson
\cite{Dyson1969,Dyson1971}. The long-range power-law model has the
advantage that by tuning the power-law exponent $\sigma$, one has access to
both the mean-field and the regimes with non-mean-field critical behavior. However, the full
power-law model is expensive for numerics. Fortunately a clever
workaround was introduced by Leuzzi et al.
\cite{PhysRevLett.101.107203} where instead of the interactions
falling off as a power law, it is the probability of there being a
bond between two spins that falls off as a power law. The fewer bonds
in the model means that a significantly smaller computational cost is
involved, thus allowing for the simulation of larger system sizes.

While the vast literature on spin glasses
is mostly focussed on Ising
spins~\cite{PhysRevLett.101.107203,PhysRevLett.103.267201,PhysRevLett.93.207203,PhysRevB.72.184416,PhysRevLett.102.177205,PhysRevB.67.134410},
there has been a revival of interest in classical $m$-component \emph{vector} spin glass
models~\cite{sharma2010almeida,sharma2011phase,sharma2011almeida,
sharma2016metastable,beyer2012one,moore:12,sharma2014avalanches,10.21468/SciPostPhys.12.1.016,lupo2018comparison,lupo2019random,baity2015soft}
in the last decade or so. The XY model has $m=2$ and the Heisenberg model has $m=3$. One of the triggers for this revival has
been the finding that the infinite-range vector spin glass exhibits an
AT line provided a magnetic field that is random in all
the component directions is applied~\cite{sharma2010almeida}. Furthermore analytical studies of the AT transition in $m$-vector models shows that the field theory of these AT transitions is that of the \emph{Ising} spin glass ~\cite{sharma2010almeida}.  Thus it has become possible to study the
question of whether or not an Ising AT transition exists in various dimensions by studying one-dimensional vector spin glasses with long-range interactions
~\cite{sharma2011almeida}!

In this paper, we study the one-dimensional diluted XY spin glass
subjected to a random vector magnetic field, with the aid of large scale
Monte Carlo simulations. While Monte Carlo simulations are a
time-tested tool for the study of phase transitions in spin glasses,
the exorbitant cost of equilibration makes them rather challenging in
practice. It has been argued that vector spins tend to equilibrate
faster compared to Ising spins~\cite{PhysRevB.80.024422}, because of
the \emph{soft} nature of the spins involved, even though the presence
of more components adds to the cost. The Heisenberg spin
glass~\cite{sharma2016metastable,sharma2011almeida,sharma2011phase,sharma2010almeida,PhysRevB.80.024422,lee2007large,PhysRevLett.97.217204,PhysRevB.34.6341,PhysRevE.61.R1008,PhysRevLett.102.027202}
has been the popular vector spin to have been considered, because of
the availability of the heatbath algorithm~\cite{lee2007large}, which
works very efficiently to equilibrate it. The XY spin glass is less
effectively handled by the heatbath
algorithm~\cite{PhysRevB.78.014419,PhysRevB.88.144104} because of the technicalities
involved in inverting a probability distribution for which a simple
closed form expression is unavailable in the XY case. In this
paper, we develop a method, which is outlined in Appendix \ref{sec:simulation details} to perform this inversion numerically
with the hope of benefiting from the vector nature of XY spins, while
simultaneously reducing the components to as small a number as
possible.

The improved algorithm yields mixed fruits. The gains from the reduced
number of components seems to be largely counterbalanced by the
additional resources consumed by the numerical inversion. However,
with the aid of extensive computational power, we are able to access
system sizes comparable to those in the corresponding study with
Heisenberg spins. Our findings for the XY diluted spin glass closely
mimic those obtained for the Heisenberg version of the same
model~\cite{sharma2011almeida,sharma2011phase,sharma2010almeida} and
the Ising spin glass in three dimensions~\cite{Baity_Jesi_2014} when
we investigate crossing the possible AT line by varying $T$ at fixed
values of $h_r$. In the mean-field regime, (we studied here the case
of $\sigma = 0.6$, which corresponds to $10$ dimensions, which is
above the upper critical dimension of $6$), there is clear evidence of
an Almeida-Thouless line (see Sec. \ref{sec:sigma_0.60}). There is
rather weak evidence for an Almeida-Thouless line for $\sigma = 0.85$
using the commonly employed finite size critical point scaling methods
of analysis (see Sec.~\ref{sec:sigma_0.85}). At this value of
$\sigma$, our system should be similar to the Edwards-Anderson model
in three dimensions with short-range interactions.  For the in-between
case at $\sigma = 0.75$ which lies in the non-mean-field regime, but
closer to the mean-field boundary at $\sigma =2/3$, our data do
provide stronger evidence for a phase transition in the presence of
small magnetic fields than at $\sigma = 0.85$. However, by varying the
magnetic field $h_r$ at fixed temperature $T$ we find in
Sec. \ref{sec:results} that at \textit{both} $\sigma = 0.75$ and at
$\sigma =0.85$ there is quite decent evidence for an AT line.
Confusingly, the field dependence of the correlation length is very
well-described by the Imry-Ma prediction of the droplet picture, which
implies the complete absence of the AT transition! In the droplet
picture the correlation length in a field remains finite and only
diverges as $h_r \to 0$. However, when this correlation length becomes
comparable to the system size $L$ the Imry-Ma formula needs to be
modified and we give in Sec. \ref{sec:dropletforms} a scaling form for
this modification. It is based upon the usual finite size scaling
approach used in studying critical phenomena, and just as for critical
phenomena we find that there are finite size corrections to this
scaling form.  In addition to these scaling corrections there are
corrections which arise when $\xi_{\text{SG}}\sim L < L^*$ which are
of different origin and are connected to TNT effects
\cite{krzakala:00, palassini:00}. TNT effects arise from droplets
whose linear dimension $L$ is of the order of the system size with
free energy cost of $O(1)$ (rather than the $L^{\theta}$ of the
droplet picture), and exist in systems whose sizes $L < L^*$
\cite{Moore:21}. The length scale $L^*$ is always large and is
expected to diverge as $d \to 6$ or as $\sigma \to 2/3$. It is only
for the case of $\sigma = 0.85$ that we can reach sizes where TNT
effects seem to be getting small.  These matters are discussed in
Sec.~\ref{TNT}.

Furthermore, we can use the droplet scaling picture to
explain some of the features of the apparent AT transition which arise
on performing the usual finite size critical scaling analyses, and
show that these are the consequence of not studying large enough
systems. Unfortunately these arguments will only become compelling for
system sizes which we cannot reach. Our chief evidence for the droplet
picture is its very successful prediction of the correlation length as
a function of the field in the region when finite size and TNT effects
are unimportant. 

Our claim that the evidence favors the absence of the AT line for
values of $\sigma$ outside the mean-field region is consistent with
the attempt~\cite{moore:12} to calculate the AT field at $T = 0$ using
an expansion in $1/m$. This indicated that as $d \to 6$ from above in
the Edwards-Anderson model, the AT field would go to zero, implying
the absence of the AT line below $6$ dimensions (which in the
one-dimensional long-range model corresponds to $2/3 < \sigma
<1$). For $\sigma > 1$ there is no finite temperature spin glass
phase.

The plan of this paper is as follows. In Sec.~\ref{sec_2} we describe
the model in detail. In Sec.~\ref{sec_3} we describe the quantities
which were studied in our Monte Carlo simulations, the details of
which are given in Appendix \ref{sec:simulation details}. Our data is
analysed in Sec.~\ref{sec:results} on the assumption that there is an
AT transition, while in Sec.~\ref{sec:dropletforms} the data is
analysed according to droplet scaling assumptions. In Sec.~\ref{TNT}
we discuss the effect of TNT behavior on our results. Finally in
Sec.~\ref{sec:discussion} we summarize our conclusions.

\section{Model Hamiltonian}\label{sec_2} 
The general Hamiltonian for vector spin glasses is:
\begin{equation}
  \mathcal{H}=-\sum\limits_{\langle i,j \rangle}J_{ij} \textbf{S}_i \cdot \textbf{S}_j
  -\sum\limits_{i}\textbf{h}_i\cdot\textbf{S}_i \,,
  \label{eqn:vector_sg_hamiltonian}
\end{equation}
where $\textbf{S}_i$ is the spin on the $i^{\text{th}}$ lattice site
($i=1,2,\ldots,N$), which is chosen to be a unit vector.  $m$
represents the number of components of the vector $\textbf{S}_i$. 
In this work we concentrate on XY spins, and set $m=2$. The
Cartesian components $h_i^{\mu}$ ($\mu=1,2$) of the on-site external
magnetic field are i.i.d random variables drawn from a Gaussian
distribution of zero mean and variance $h_r^2$ and satisfy the
relation:
\begin{equation}
  \left[ h_i^{\mu} h_j^{\nu} \right]_{\text{av}} = h_r^2\delta_{ij} \delta_{\mu \nu}.
  \label{eqn:h-correlation}
\end{equation}
We use the notation $\langle\cdots\rangle$ for thermal average and
$[\cdots]_{\text{av}}$ for an average over quenched disorder
throughout this paper.

The spins are arranged on a circle so the
geometric distance between a pair of spins $(i,j)$ is given
by~\cite{PhysRevB.67.134410} 
\begin{equation}
  r_{ij}=\frac{N}{\pi}\sin\left(\frac{\pi}{N}\left| i-j \right| \right),
  \label{eqn:distance}
\end{equation}
which is the length of the chord connecting the $i^{\text{th}}$ and
$j^{\text{th}}$ spins. The
interactions $J_{ij}$ are independent random variables such that the
probability of having a non-zero interaction between a pair of spins
$(i,j)$ falls with the distance $r_{ij}$ between the spins as a power
law:
\begin{equation}
P(J_{ij}) \propto \frac{1}{r_{ij}^{2\sigma}}.  
\end{equation}
If the spins $i$ and $j$ are linked the magnitude of the interaction
between them is drawn from a Gaussian distribution whose mean is zero
and whose standard deviation is unity, i.e:
\begin{equation}
  \left[ J_{ij} \right]_{\text{av}}=0 \qquad \text{and} \qquad \left[ J_{ij}^2 \right]_{\text{av}} = J^2 = 1.
  \label{eqn:vb-interaction}
\end{equation}
The mean number of non-zero bonds from a site is fixed to be $\tilde{z}$
(co-ordination number). So, the total number of bonds among all the
spins on the lattice is fixed to be $N_b=N \tilde{z} /2$. When $\tilde{z}=6$ this model
mimics the 3D simple cubic lattice model and we use this value for $\tilde{z}$
for all the $\sigma$ values studied.  (For $\sigma=0$ and
$\tilde{z}=N-1$, the model becomes the infinite-range Sherrington-Kirkpatrick
(SK) model \cite{sherrington:75}).

To generate the set of interaction
pairs~\cite{PhysRevLett.101.107203,sharma2011phase} $(i,j)$ with the
desired probability we pick a site $i$ randomly and uniformly and then
choose a second site $j$ with probability given by:
\begin{equation}
  p_{ij}=\frac{r_{ij}^{-2\sigma}}{\sum\limits_{j\neq i}r_{ij}^{-2\sigma}}.
  \label{eqn:probability}
\end{equation}
If the spins at $i$ and $j$ are already connected we repeat this process
until we find a pair of sites $(i,j)$ which have not been
connected. Once we find such a pair of spins, we connect them with a
bond whose strength $J_{ij}$ is a Gaussian random variable with
attributes given by Eq. (\ref{eqn:vb-interaction}).  We repeat this
process exactly $N_b$ times to generate $N_b$ pairs of interacting
spins.

The advantage of the diluted model over the fully connected model is
that, in a fully connected model, there are $N(N-1)/2$
interactions. The ratio of the number of interactions of the diluted
model to the fully connected model is $\tilde{z}/N$ which is a very small
value as $N$ becomes large. Hence it is possible to go to much larger
system sizes with a diluted model as compared to a fully connected
model.

\begin{figure*}[ht]
  \centering
  \includegraphics[width=\textwidth]{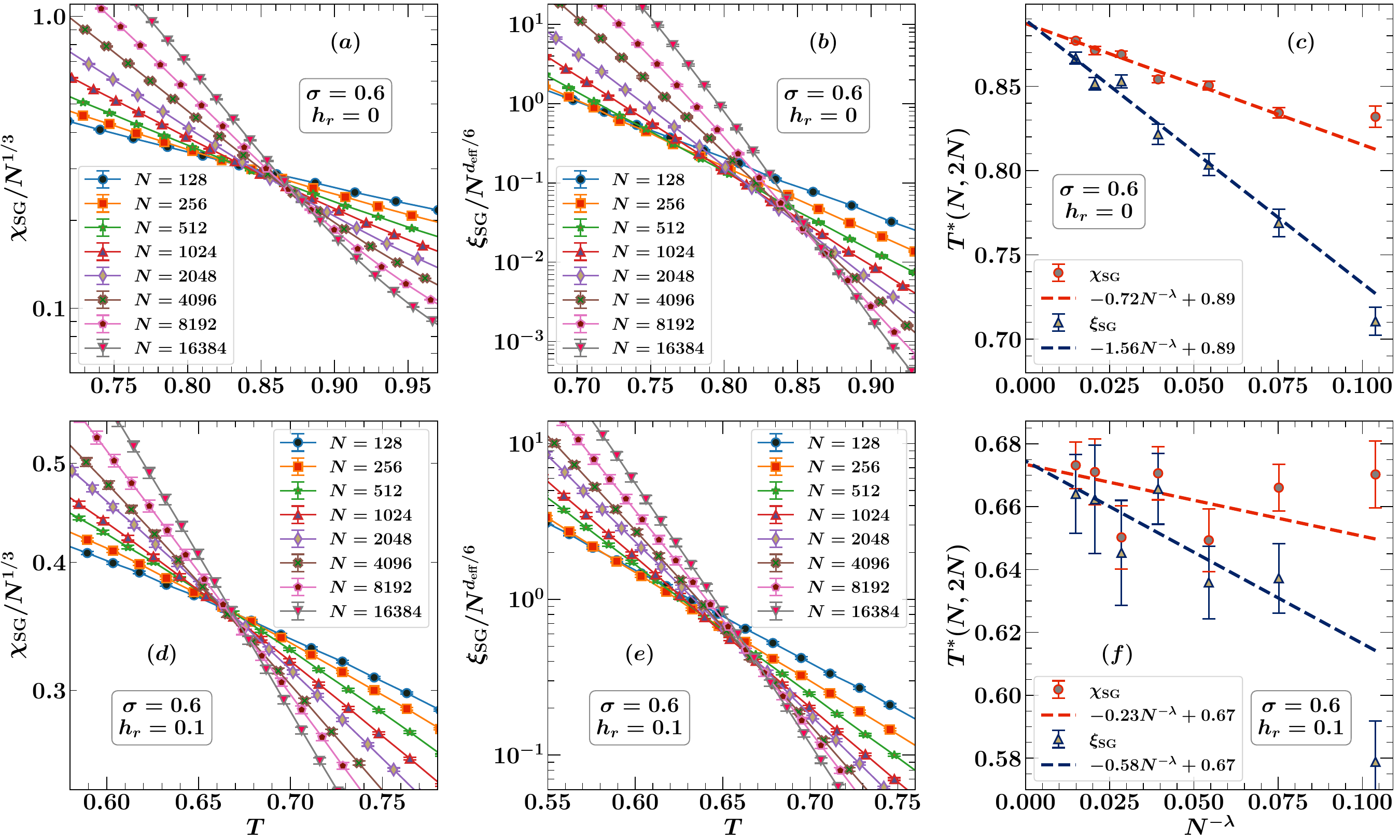}
  \caption{Finite size scaling analyses of data for $\sigma=0.60$
      obtained by fixing the field and varying the temperature. The
      zero-field data are presented in the first row and the data with
      $h_r=0.1$ are shown in the second row. Figures (a) and (d) show
      the plots of $\chi_{\text{SG}}/N^{1/3}$ as a function of the
      temperature $T$ for different system sizes, with $h_r=0$ and
      $h_r=0.1$ respectively. The corresponding data for
      $\xi_{\text{SG}}/N^{d_{\text{eff}}/6}$ are shown in figures (b)
      and (e), with $d_{\text{eff}}=2/(2\sigma-1)$ in the mean-field
      regime (Eq.~(\ref{eq:d_eff})). The exponents of $N$ are chosen
      according to Eqs.~(\ref{eq:chi_fss:1/2<sigma<2/3}) and
      (\ref{eq:xi_fss:1/2<sigma<2/3}). Both the sets of plots show
      that the curves for different system sizes intersect. The data
      for the intersection temperatures $T^*(N,2N)$ between pairs of
      adjacent system sizes for $\chi_{\text{SG}}/N^{1/3}$ and
      $\xi_{\text{SG}}/N^{d_{\text{eff}}/6}$ are plotted as a function
      of $N^{-\lambda}$ in figures (c) and (f), for $h_r=0$ and
      $h_r=0.1$ respectively. The value of the exponent $\lambda$ is
      fixed to be $0.467$ which is known exactly in the mean-field
      regime~\cite{sharma2011phase,PhysRevB.81.064415}. For $h_r=0$,
      the fits give $T_c=0.8873 \pm 0.0017$ from $\chi_{\text{SG}}$
      and $T_c=0.8893 \pm 0.0046$ from $\xi_{\text{SG}}$. For
      $h_r=0.1$, both the datasets are consistent with a spin glass
      transition temperature of $T_{\text{AT}}(h_r=0.1) = 0.67$.}
  \label{fig:chi_cl_vs_T_s0.60}
\end{figure*}

At zero-field, the mean-field spin glass transition temperature for
the $m$-component vector spin glass is given
by~\cite{Almeida_1978,sharma2011phase,sharma2011almeida}
\begin{equation}
  \label{eq:T_c^MF_zero_field}
  T_c^{\text{MF}}=\frac{1}{m} \left( \sum_j \left[J_{ij}^2 \right]_{\text{av}} \right)^{1/2}
  =\frac{\sqrt{\tilde{z}}}{m} J.
\end{equation}
The approximate location of the AT line for an $m$-component infinite-range spin
glass near the zero-field transition temperature $T_c$ is~\cite{sharma2010almeida}
\begin{equation}
 \label{eq:approx_AT_line}
 \left( \frac{h_r}{J} \right)^2 = \frac{4}{m(m+2)} \left( 1-\frac{T}{T_c}
  \right)^3.
\end{equation}
The accuracy of this approximation for the SK model can be judged from Fig.~\ref{fig:XY_AT}.

A one-dimensional chain with power law diluted interactions for a
particular value of $\sigma$ is equivalent to a short-range
model~\cite{PhysRevLett.102.177205,PhysRevB.81.064415} of effective
dimension $d_{\text{eff}}$, where
\begin{equation}
  \label{eq:d_eff}
  d_{\text{eff}}=\frac{2}{2\sigma-1},
\end{equation}
i.e., there is a one-to-one mapping between a long-range diluted
network with exponent $\sigma$ and a short-range model with space
dimension $d_{\text{eff}}$, at least when $1/2 <\sigma<2/3$. Thus when
$\sigma = 0.60$, $d_{\text{eff}}=10$. For the interval
$2/3 <\sigma <1$ other relations are required
~\cite{PhysRevLett.102.177205, Banos:12}. For example, for Ising spin
glasses, it was suggested in Ref. ~\cite{Banos:12} that $d= 4$
corresponded to $\sigma\approx 0.790$, while $d=3$ corresponded to
$\sigma \approx 0.896$. Unfortunately, the mapping for the XY model
has been less studied.

\begin{figure*}
  \centering
  \includegraphics[width=\textwidth]{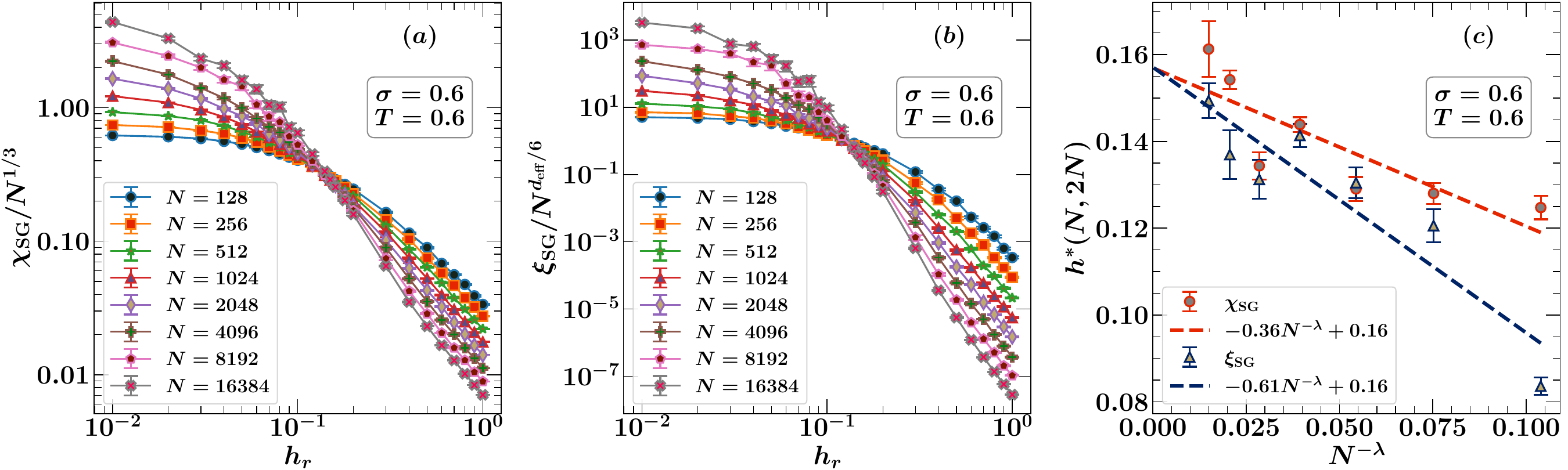}
  \caption{Finite size scaling analyses of $\chi_{\text{SG}}$ data
    (figure (a)), and $\xi_{\text{SG}}$ data (figure (b)), for
    $\sigma=0.60$ obtained by fixing the temperature to $T=0.6$
    $(=0.67\,T_c)$ and varying the field. Both the plots show that the
    curves for different system sizes intersect. Figure (c) shows the
    data for the intersection fields $h^*(N,2N)$ between pairs of
    adjacent system sizes, plotted as a function of
    $N^{-\lambda}$. Both the datasets are consistent with a spin glass
    transition at $h_{\text{AT}}(T=0.6) \approx 0.16$.}
  \label{fig:chi_cl_vs_hr_s0.60}
\end{figure*}

\section{Correlation lengths and susceptibilities}
\label{sec_3}

In this section we discuss the quantities which were obtained from our
Monte Carlo simulations and used to extract a correlation length
$\xi_{\text{SG}}$ and the spin glass susceptibility
$\chi_{\text{SG}}$. The simulations themselves are described in detail
in Appendix \ref{sec:simulation details}.

The thermal average of a quantity is calculated using multiple
replicas in the following standard way:
\begin{equation}
  \label{eq:thermal_average_replicas}
  \langle A \rangle \langle B \rangle \langle C \rangle \langle D \rangle = \langle A^{(1)} B^{(2)} C^{(3)} D^{(4)} \rangle
\end{equation}
where (1),(2),(3), and (4) are four copies of the system at the same
temperature, and $A$,$B$,$C$, and $D$ are the quantities over which we
would like to perform thermal averaging. The wave-vector-dependent
spin glass susceptibility is given by~\cite{sharma2010almeida}
\begin{equation}
  \label{eq:wvd_sg_susceptibility}
  \chi_{\text{SG}}(k)=\frac{1}{N}\sum\limits_{i,j}\frac{1}{m}\sum\limits_{\mu,\nu}
  \left[\left(\chi_{ij}^{\mu\nu} \right)^2 \right]_{\text{av}}e^{ik(i-j)} ,
\end{equation}
where
\begin{equation}
  \label{eq:chi_ij^mu-nu}
  \chi_{ij}^{\mu\nu}=\left\langle S_i^{\mu} S_j^{\nu} \right\rangle -
  \left\langle S_i^{\mu} \right\rangle\left\langle S_j^{\nu} \right\rangle.
\end{equation}
The spin glass correlation length is then determined from
\begin{equation}
  \label{eq:sg_correlation_length}
  \xi_{\text{SG}}=\frac{1}{2\sin(k_{\text{min}}/2)}
  \left(\frac{\chi_{\text{SG}}(0)}{\chi_{\text{SG}}\left( k_{\text{min}} \right)} -1 \right)^{1/(2\sigma-1)}
\end{equation}
where $k_{\text{min}}=(2\pi/N)$.

\begin{figure*}
  \centering
  \includegraphics[width=\textwidth]{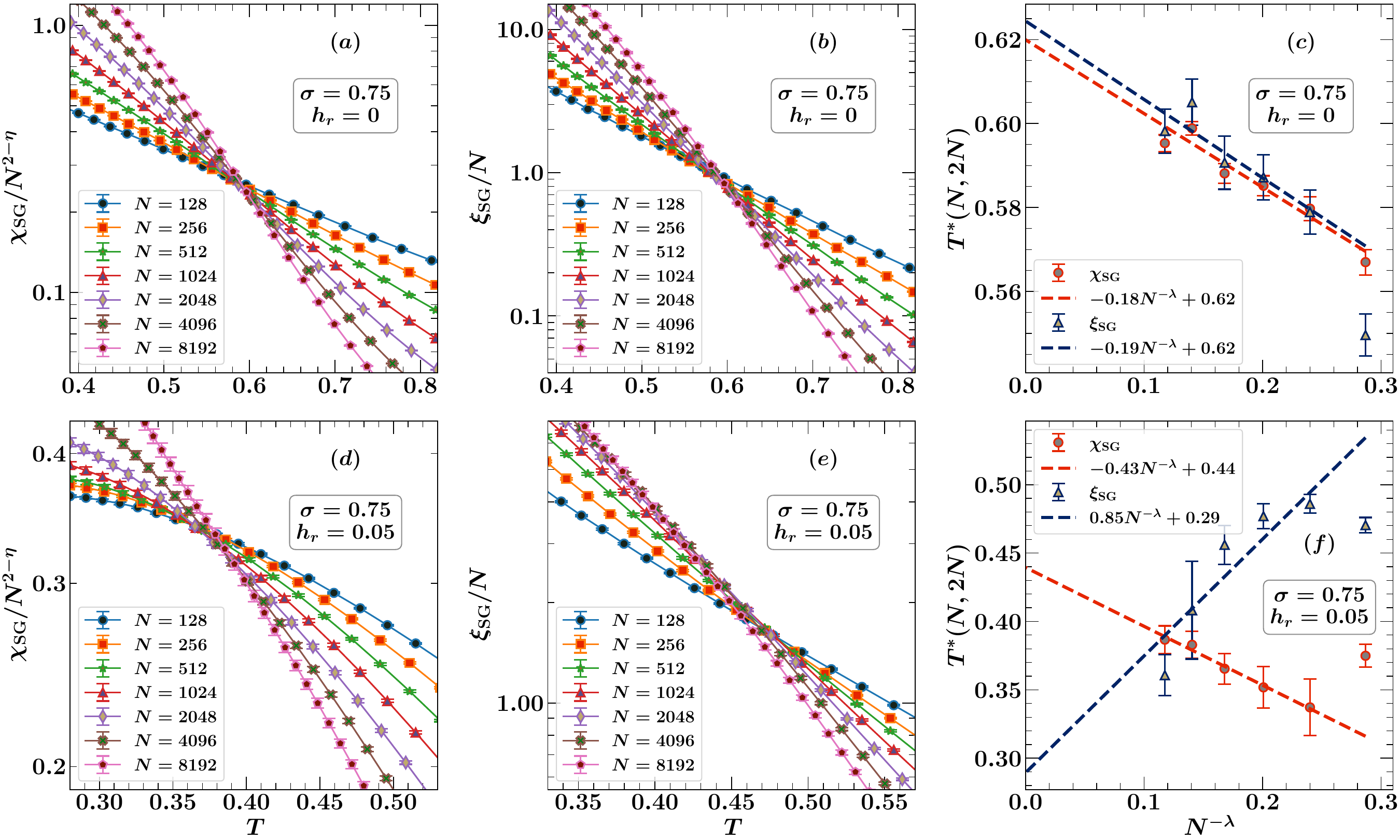}
  \caption{Finite size scaling analyses of data for $\sigma = 0.75$
    obtained by varying the temperature at fixed field. Figures in
    first row show the data generated at zero-field, where as the
    figures in second row show the data obtained by turning on a
    magnetic field of $h_r=0.05$. Figures (a) and (d) show data for
    $\chi_{\text{SG}}/N^{2-\eta}$ (with $2-\eta=2\sigma-1$) for
    different system sizes. Figures (b) and (e) show the corresponding
    data for $\xi_{\text{SG}}/N$. According to
    Eqs.~(\ref{eq:chi_fss:2/3<sigma<1}) and
    (\ref{eq:xi_fss:2/3<sigma<1}) the data should intersect at
    $T_{\text{AT}}(h_r)$, which is shown in figures (c) and (f) for
    $h_r=0$ and $h_r=0.05$ respectively. The $T^*(N,2N)$ data shown in
    these figures did not fit well with
    Eq.~(\ref{eq:intersection_temperature_corrections}). So we used
    the value of the scaling exponent $\lambda=0.26$ obtained from the
    $h^*(N,2N)$ data (see Table~\ref{tab:lambda}), and fitted the
    $T^*(N,2N)$ data against $N^{-\lambda}$ using a straight line. The
    resulting values for the zero-field transition temperature are
    $T_{c} = 0.6200 \pm 0.0031$ from $\chi_{\text{SG}}$ and
    $T_{c} = 0.6244 \pm 0.0098$ from $\xi_{\text{SG}}$. For $h_r=0.05$
    the linear fitting gives
    $T_{\text{AT}}(h_r=0.05) = 0.4395 \pm 0.0241$ from
    $\chi_{\text{SG}}$ and
    $T_{\text{AT}}(h_r=0.05) = 0.2893 \pm 0.0252$ from
    $\xi_{\text{SG}}$, and the values do not agree with each other.}
  \label{fig:chi_cl_vs_T_s0.75}
\end{figure*}

\section{Finite-size analyses assuming a transition exists}
\label{sec:ATforms}

In this section we detail the method of finite-size analysis when a
transition is assumed to exist. When studying the AT line, which is a
line of phase transitions in the $h_r-T$ plane, it can be crossed on
an infinite number of trajectories. The most commonly used trajectory
is the one where $h_r$ is kept constant and the temperature $T$ is
varied. In this work we also consider the trajectory in which $T$ is
kept constant and $h_r$ is varied. We refer to the zero-field
transition temperature as $T_c$ while we denote a generic transition
temperature on the AT line by $T_{\text{AT}}(h_r)$. Similarly we
denote the field on the AT line by $h_{\text{AT}}(T)$.

The spin glass susceptibility
$\chi_{\text{SG}} \equiv \chi_{\text{SG}}(0)$ of a finite system of
$N$ spins has the finite size scaling form (near the transition
temperature $T_{\text{AT}}(h_r)$)~\cite{sharma2010almeida}:
\begin{subequations}
  \label{eq:chi_finite-system}
  \begin{align}
    \label{eq:chi_fss:2/3<sigma<1}
    \frac{\chi_{\text{SG}}}{N^{2-\eta}} &= \mathcal{C}\left[ N^{1/\nu}\left(T-T_{\text{AT}}(h_r)\right)\right],\quad
                                          (2/3 \leq \sigma < 1), \\
    \label{eq:chi_fss:1/2<sigma<2/3}
    \frac{\chi_{\text{SG}}}{N^{1/3}} &= \mathcal{C}\left[ N^{1/3}\left(T-T_{\text{AT}}(h_r)\right)\right],\quad
                                       (1/2 < \sigma \leq 2/3),
  \end{align}
\end{subequations}
where $\eta$ is given by $2-\eta=2\sigma-1$.  These forms are examples
of finite size scaling expressions which would be expected to hold in
the critical region when $N \to \infty$, $(T-T_{\text{AT}}(h_r)) \to 0$, with (say)
$N^{1/\nu}(T-T_{\text{AT}}(h_r))$ finite. The scaling function $\mathcal{C}$ will
depend on the value of $\sigma$. There are always finite size
corrections to these forms. For example, the corrections to
Eq.~(\ref{eq:chi_fss:1/2<sigma<2/3}) will be of the form
\begin{eqnarray}
 \frac{\chi_{\text{SG}}}{N^{1/3}} &=& \mathcal{C}\left[ N^{1/3}\left(T-T_{\text{AT}}(h_r)\right)\right]\nonumber\\
 &+&N^{-\omega} \mathcal{G}\left[N^{1/3}(T-T_{\text{AT}}(h_r))\right].
 \label{q:chi_fss:1/2<sigma<2/3corr}
\end{eqnarray}
It has been suggested \cite{sharma2011phase,PhysRevB.81.064415} that
the correction to scaling exponent is given at least in the mean-field
region by
\begin{equation}
\omega=1/3-(2\sigma-1).
\label{eq:omega?}
\end{equation}
Curves of $\chi_{\text{SG}}/N^{2-\eta}$ ($\chi_{\text{SG}}/N^{1/3}$ in
the mean-field regime) plotted for different system sizes should
intersect at the transition temperature $T_{\text{AT}}(h_r)$. In reality, finite-size
corrections to Eq.~(\ref{eq:chi_finite-system}) are always present and
cause the intersection point between the curves for size $N$ and $2N$
to depend on $N$.  The intersection temperatures vary
as~\cite{PhysRevB.81.064415,Binder1981,BALLESTEROS1996125,PhysRevB.78.214205}
\begin{equation}
  \label{eq:intersection_temperature_corrections}
  T^*(N,2N) = T_{\text{AT}}(h_r) + \frac{A}{N^{\lambda}} ,
\end{equation}
where $A$ is the amplitude of the leading correction, and the exponent
$\lambda$ is
\begin{subequations}
\label{eq:lambdaval}
\begin{align}
  \label{eq:lambda_mean-field} 
  \lambda &=1/3+\omega,\quad                                     (1/2 < \sigma \leq 2/3),\\
  \label{eq:lambda_nonmeanfield}
  \lambda &=1/\nu +\omega,\quad         
                                        (2/3 < \sigma < 1),                                       
\end{align}  
\end{subequations}
where $\omega$ is the leading correction to the scaling exponent. When
$\sigma = 0.6$, $\omega= -2 \sigma +4/3$, so
$\lambda=5/3-2 \sigma=0.467$ \cite{PhysRevB.81.064415}. In the regime
when $\sigma> 2/3$ the values of both $\nu$ and $\lambda$ are not
well-determined, so there we shall treat $\lambda$ as a fitting
parameter.

The spin glass correlation
length has a similar finite size scaling form in the critical region
\begin{subequations}
  \label{eq:xi_finite-system}
  \begin{align}
    \label{eq:xi_fss:2/3<sigma<1}
    \frac{\xi_{\text{SG}}}{N} &= \mathcal{X}\left[ N^{1/\nu}\left(T-T_{\text{AT}}(h_r)\right)\right],\quad
                    (2/3 \leq \sigma < 1), \\
    \label{eq:xi_fss:1/2<sigma<2/3}
    \frac{\xi_{\text{SG}}}{N^{d_{\text{eff}}/6}} &= \mathcal{X}\left[ N^{1/3}\left(T-T_{\text{AT}}(h_r)\right)\right],\quad
                            (1/2 < \sigma \leq 2/3).
  \end{align}
\end{subequations}
$\nu$, the correlation length critical exponent, has to be determined
numerically in the interval $2/3 < \sigma < 1$.

\begin{figure*}
  \centering
  \includegraphics[width=\textwidth]{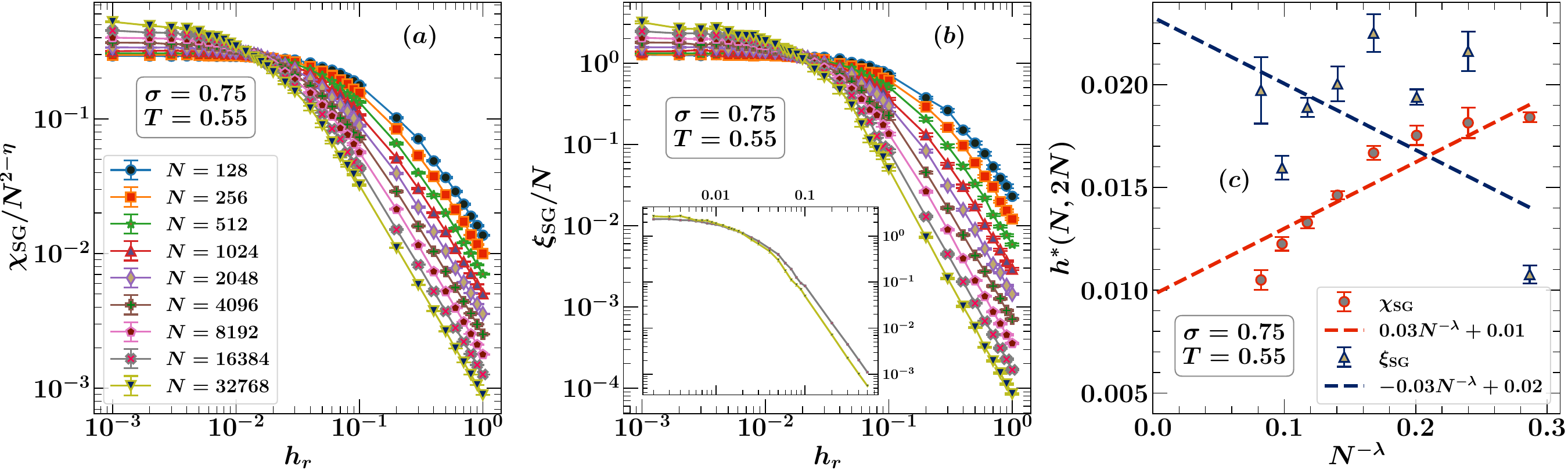}
  \caption{Finite size scaling analyses of data assuming a transition
    exists, for $\sigma=0.75$ at a temperature of $T=0.55$
    $(=0.89\,T_c)$, are shown: (a) for $\chi_{\text{SG}}$ and (b) for
    $\xi_{\text{SG}}$ (inset shows our two largest system sizes
    $N=16384$ and $N=32768$). The legend displayed in figure (a) is
    common for both the figures (a) and (b). A non-linear fit of the
    intersection fields data obtained from $\chi_{\text{SG}}$ (shown
    in figure (a)), with the
    Eq.~(\ref{eq:intersection_temperature_corrections}) using the
    Levenberg-Marquadt algorithm gives $\lambda=0.26$ (see Table
    \ref{tab:lambda}). Figure (c) shows the $h^*(N,2N)$ data fitted
    against $N^{-\lambda}$ with a straight line.}
  \label{fig:chi_cl_vs_hr_s0.75}
\end{figure*}

We have also studied crossing the AT line at fixed $T$ and varying
$h_r$. Then Eq.~(\ref{eq:xi_finite-system}) takes the form,
\begin{subequations}
  \label{eq:xi_finite-systemh}
  \begin{align}
    \label{eq:xi_fss:2/3<sigma<1h}
    \frac{\xi_{\text{SG}}}{N} &= \mathcal{X}\left[ N^{1/\nu}\left(h_r-h_{\text{AT}}(T)\right)\right],\quad
                    (2/3 \leq \sigma < 1), \\
    \label{eq:xi_fss:1/2<sigma<2/3h}
    \frac{\xi_{\text{SG}}}{N^{d_{\text{eff}}/6}} &= \mathcal{X}\left[ N^{1/3}\left(h_r-h_{\text{AT}}(T)\right)\right],\quad
                            (1/2 < \sigma \leq 2/3),
  \end{align}
\end{subequations}
where $h_{\text{AT}}(T)$ denotes the field at the AT line at temperature $T$. Similarly, the spin glass susceptibility $\chi_{\text{SG}}$ of the finite system near the AT transition line takes the form
\begin{subequations}
  \label{eq:chi_finite-systemh}
  \begin{align}
    \label{eq:chi_fss:2/3<sigma<1h}
    \frac{\chi_{\text{SG}}}{N^{2-\eta}} &= \mathcal{C}\left[ N^{1/\nu}\left(h_r-h_{\text{AT}}(T)\right)\right],\quad
                    (2/3 \leq \sigma < 1), \\
    \label{eq:chi_fss:1/2<sigma<2/3h}
    \frac{\chi_{\text{SG}}}{N^{1/3}} &= \mathcal{C}\left[ N^{1/3}\left(h_r-h_{\text{AT}}(T)\right)\right],\quad
                            (1/2 < \sigma \leq 2/3).
  \end{align}
\end{subequations}

In the thermodynamic limit, Eq.~(\ref{eq:xi_finite-system}) is similar
to Eq.~(\ref{eq:xi_finite-systemh}); the effect of finite size
corrections to the two can differ. For example, while the correction
to scaling exponent $\lambda$ does not depend on the choice of the
trajectory, the magnitude of the scaling corrections can differ. Thus
in the intersection formulae when applied to fields
\begin{equation}
  \label{eq:intersection_field_corrections}
  h^*(N,2N) = h_{\text{AT}}(T) + \frac{\tilde{A}}{N^{\lambda}},
\end{equation}
the coefficient $\tilde{A}$ will be different from $A$ in
Eq.~(\ref{eq:intersection_temperature_corrections}). Corrections to
scaling of, say, Eq.~(\ref{eq:chi_fss:2/3<sigma<1h}), are more
generally of the form
\begin{eqnarray}
\frac{\chi_{\text{SG}}}{N^{2-\eta}}&=&\mathcal{C}\left[ N^{1/\nu}\left(h_r-h_{\text{AT}}(T)\right)\right] \nonumber \\ &+& N^{-\omega}\mathcal{G}\left[ N^{1/\nu}\left(h_r-h_{\text{AT}}(T)\right)\right],
\label{eq:scalingcorrform1}
\end{eqnarray}
where $\omega$ is the correction to scaling exponent, and
$\mathcal{G}$ is another scaling function. This type of scaling form
holds in the limit where $N^{1/\nu}(h_r-h_{\text{AT}}(T))$ is fixed as
$N\to \infty$, which of course can only be realized approximately in
numerical studies.

A key feature of the finite size critical point scaling analysis is
that right on the AT line itself, that is when $h_r=h_{\text{AT}}(T)$,
$R = \chi_{\text{SG}}/ N^{2-\eta}$ ($\chi_{\text{SG}}/ N^{1/3}$ for
$\sigma\leq 2/3$) should be finite as $N\to \infty$. We find (see
Sec.~\ref{sec:dropletforms}) that $R$ is at least not increasing with
$N$, and perhaps finite (see Fig. \ref{fig:Rxlog}\flc{a}), for
$\sigma= 0.60$ but for $\sigma = 0.70,\, 0.75$ and $0.85$ it is in
fact increasing with $N$, at the crossing field $h^*(N, 2N)$. We
deduce from this observation that at these values of $\sigma$ the
crossings at $h^*(N, 2N)$ are not associated with a true critical
point at all but are consequences of droplet scaling.  At a true
critical point $R$ would tend to a finite constant as $N$ increases,
but we find it increases with $N$, provided $N >1024$ (or system sizes
$2N> 2048$) for the case of $\sigma = 0.75$ (see
Fig. \ref{fig:Rxlog}\flc{b}).

In Sec. \ref{sec:results} we shall present our attempts at analysing
the data for $\sigma= 0.6$, $\sigma = 0.75$, $\sigma = 0.85$ at fixed
values of $h_r$ but varying $T$, and also at a fixed value of $T$ and
varying $h_r$, on the assumption that there \emph{is} an AT line and
using the finite-size scaling methods of this subsection.

We have also obtained data at fixed $T$ and varying $h_r$ for
$\sigma =0.60,\,\, 0.70,\,\, 0.75,\,\,0.85$ and analysed them using
finite size generalizations of well-known droplet scaling
relations. In this case the droplet picture provides a simple set of
formulae for analysing the data in the assumed \emph{absence} of an AT
line.

\begin{figure*}
  \centering
  \includegraphics[width=\textwidth]{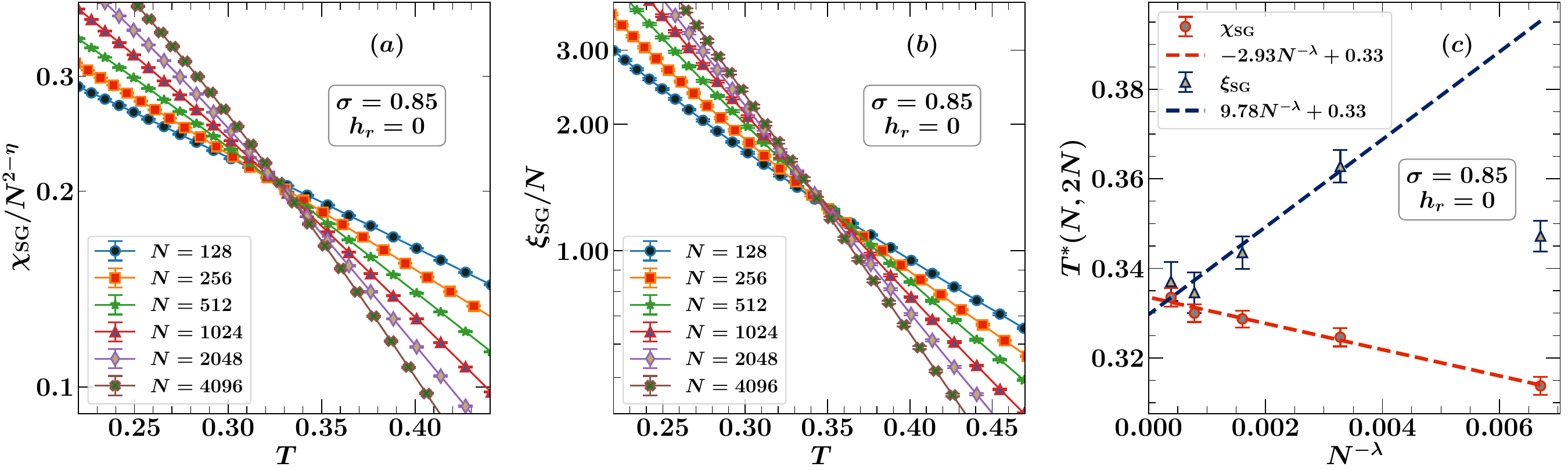}
  \caption{A finite size scaling plot of $\chi_{\text{SG}}$ (figure
    (a)) and $\xi_{\text{SG}}$ (figure (b)), for $\sigma=0.85$ in the
    absence of magnetic field ($h_r=0$) (with
    $2-\eta=2\sigma-1$). Both the datasets clearly indicate that a
    phase transition occurs. The transition temperature in the
    thermodynamic limit is estimated in figure (c). A non-linear fit
    of the $\chi_{\text{SG}}$ data from figure (c) with the
    Eq.~(\ref{eq:intersection_temperature_corrections}) using the
    Levenberg-Marquadt algorithm gives $\lambda=1.03$ (see
    Table~\ref{tab:lambda}). A linear fit of the data using this value
    of $\lambda$ gives $T_{c} = 0.3336 \pm 0.0013$ from
    $\chi_{\text{SG}}$ and $T_{c} = 0.3297 \pm 0.0036$ from
    $\xi_{\text{SG}}$.}
  \label{fig:chi_cl_vs_T_s0.85}
\end{figure*}

\section{Analyses of the simulation data assuming there is an AT line} 
\label{sec:results}

We shall study the phase transitions at $h_r=0$, and determine the
zero-field transition temperature $T_c$ ($=T_{\text{AT}}(h_r=0)$), and seek evidence of an AT
transition at non-zero $h_r$ using the standard critical point finite
size scaling method of determining the ``crossings'' or intersections
of the curves of, say, $\chi_{\text{SG}}/N^z$ (with $z =1/3$ when
$\sigma \leq 2/3$, and with $z = 2-\eta= 2\sigma-1$ for
$\sigma \geq 2/3$) at values of $N$ and $2N$ as we reduce $T$ through
the AT transition temperature at fixed $h_r$, or the field $h_r$ at
fixed $T$ in the vicinity of the AT field $h_{\text{AT}}(T)$ as outlined in
Sec.~\ref{sec:ATforms}. There seems no reason to doubt the existence
of an AT line for any value of $\sigma$ in the mean-field region
$\sigma < 2/3$, and our results are entirely consistent with the
existence of an AT transition at $\sigma = 0.60$. They serve as a
useful comparison for the studies in the non-mean-field regime
$\sigma > 2/3$, where the evidence will be found to favor the droplet
picture.  We have studied $N$ values 128, 256, 512, 1024, 2048, 4096,
8192 and 16384 for both $\sigma = 0.60$ and $\sigma = 0.75$, but went
up to $N=32768$ for the case of $\sigma = 0.75$ when the field $h_r$
was varied at fixed $T$. When $\sigma = 0.85$ the largest $N$ values
used was $4096$. In this case the zero-field transition temperature
$T_c$ is quite low and as a consequence all the investigations
have to be done also at low temperatures, where equilibration times
are long, preventing the study of larger systems.  We are mainly
interested in the question as to whether outside the mean-field
region, that is for $\sigma >2/3$, an AT transition actually exists
and whether (say) the dependence of $\xi_{\text{SG}}$ on the field
$h_r$ can be understood as will be suggested in
Sec. \ref{sec:dropletforms} on the droplet picture without invoking an
AT transition at all. If it can, this would provide support to the
argument that the droplet scaling picture rather than replica symmetry
breaking describes spin glasses below $6$ dimensions. We have analysed
the data for $\sigma =0.70,\,\, 0.75$ and for $\sigma =0.85$ using the
usual ``crossing'' method, (the finite size scaling approach outlined
in Sec.~\ref{sec:ATforms}), which indeed works well for
$\sigma = 0.6$.  The evidence for the existence of an AT transition at
$\sigma = 0.75$ and $0.85$ will be contrasted with the evidence
against an AT transition at these values of $\sigma$.

Our main focus was the case $\sigma = 0.75$. We looked briefly at the
case $\sigma = 0.70$ to find whether or not it might be practical to
study whether $\sigma = 2/3$ is the value of $\sigma$ above which the
AT line might disappear. We found that it was similar to
$\sigma = 0.75$, but that the corrections to scaling were larger. This
means that for a given level of accuracy, larger $N$ values are
required. We studied $\sigma = 0.85$ because it should behave
similarly to physical systems in three dimensions but we could not
equilibrate systems at the larger $N$ values in this case because the
temperatures $T$ of interest have to be less than $T_c$, which is
rather small.
 
\subsection{$\sigma=0.6$}
\label{sec:sigma_0.60}

\begin{figure*}[t]
  \centering
  \includegraphics[width=\textwidth]{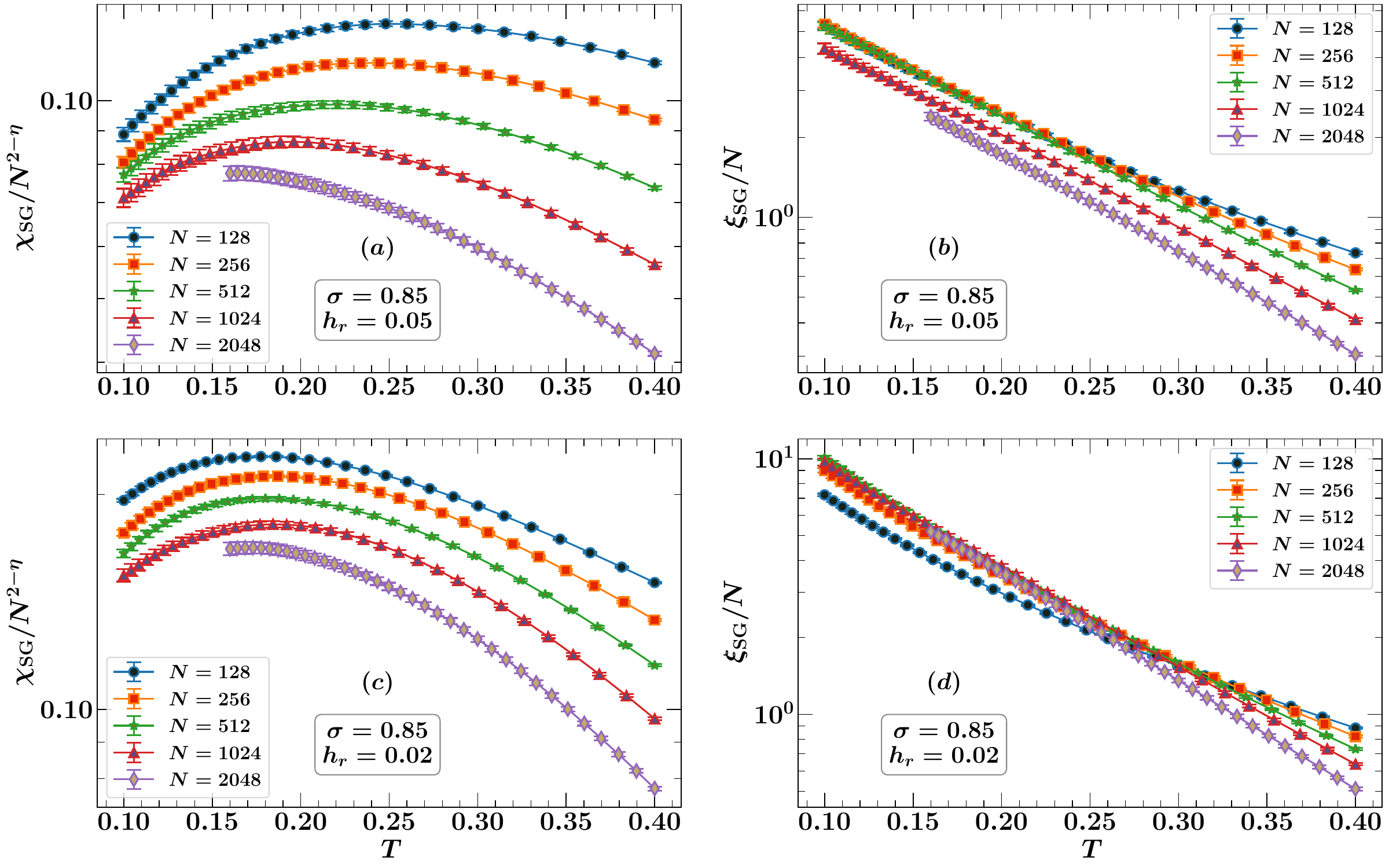}
  \caption{Finite size scaling plots for $\sigma=0.85$ are shown: (a)
    for $\chi_{\text{SG}}$ with $h_r=0.05$, (b) for $\xi_{\text{SG}}$
    with $h_r=0.05$, (c) for $\chi_{\text{SG}}$ with $h_r=0.02$, and
    (d) for $\xi_{\text{SG}}$ with $h_r=0.02$. Figure (c) shows that
    the data do not intersect even at very low temperatures (much
    lower than the mean field value of
    $T_{\text{AT}}(h_r=0.02) = 0.2997$ obtained using
    Eq.~(\ref{eq:approx_AT_line})) indicating that there is no phase
    transition in this regime. In figures (b) and (d), the data show
    merging behavior at low temperatures.}
  \label{fig:chi_cl_vs_T_s0.85_infield}
\end{figure*}

We shall focus on $\sigma =0.60$ in this subsection. It corresponds
according to Eq.~(\ref{eq:d_eff}) to an effective dimension of $10$
dimensions, which is in the mean-field region; it lies above the upper
critical dimension of spin glasses, which is $6$ (or in the mean-field
region $\sigma < 2/3$ in the one-dimensional long-range model). It is
natural to expect that for this value of $\sigma$ there will be an AT
line and this is amply confirmed by our simulations. For this value of
$\sigma$, simulations of the corresponding Ising
model~\cite{PhysRevLett.103.267201,PhysRevLett.102.177205} and the
Heisenberg model~\cite{sharma2011almeida,sharma2011phase} also found
an AT line.

Our results for $h_r=0$ are given in
Figs.~\ref{fig:chi_cl_vs_T_s0.60}\flc{a},
\ref{fig:chi_cl_vs_T_s0.60}\flc{b}, and
\ref{fig:chi_cl_vs_T_s0.60}\flc{c}. According to
Eq. (\ref{eq:chi_fss:1/2<sigma<2/3}), the data for
$\chi_{\text{SG}}/N^{1/3}$ when plotted for different system sizes
should intersect at the transition temperature $T_c$. Similarly,
according to Eq. (\ref{eq:xi_fss:1/2<sigma<2/3}), the data of
$\xi_{\text{SG}}/N^{d_{\text{eff}}/6}$ with
$d_{\text{eff}}=2/(2\sigma-1)$ should intersect at the same transition
temperature. Figs. \ref{fig:chi_cl_vs_T_s0.60}\flc{a} and
\ref{fig:chi_cl_vs_T_s0.60}\flc{b} show the data for different
system sizes.  We find the temperature $T^*(N,2N)$ at which the curves
corresponding to the system sizes $N$ and $2N$ intersect. We then fit
this data with Eq. (\ref{eq:intersection_temperature_corrections}) to
find the transition temperature.  The exponent
$\lambda \equiv 5/3-2\sigma$ is known to equal $0.467$ in this case
~\cite{sharma2011phase,PhysRevB.81.064415}.  The result is displayed
in Fig.~\ref{fig:chi_cl_vs_T_s0.60}\flc{c}, where the $T^*(N,2N)$
data obtained from intersections of $\chi_{\text{SG}}$ are fitted
against $N^{-\lambda}$ with a straight line for the largest $6$ pairs
of system sizes to give $T_{c}= 0.8873\pm 0.0017$.  The corresponding
intersections of the $\xi_{\text{SG}}$ data (omitting the two smallest
system sizes) gives $T_{c}=0.8893\pm 0.0046$.  The values of $T_{c}$
obtained from $\chi_{\text{SG}}$ data and $\xi_{\text{SG}}$ data are
in agreement with each other. The mean-field prediction of
Eq.~(\ref{eq:T_c^MF_zero_field}) is much higher,
$T_c^{\text{MF}}=\sqrt{6}/2=1.2247$. Fluctuation effects not present
in the SK limit must be responsible for this large difference.

For $h_r=0.1$, the data is as shown in
Figs.~\ref{fig:chi_cl_vs_T_s0.60}\flc{d},
\ref{fig:chi_cl_vs_T_s0.60}\flc{e}, and
\ref{fig:chi_cl_vs_T_s0.60}\flc{f}. When the $T^*(N,2N)$ data obtained
from $\chi_{\text{SG}}$ are fitted against $N^{-\lambda}$ with a
straight line for the largest $4$ pairs of system sizes we get
$T_{\text{AT}}(h_r=0.1) = 0.6735\pm 0.0120$. The corresponding
$\xi_{\text{SG}}$ data (omitting the two smallest system sizes) gives
$T_{\text{AT}}(h_r=0.1) = 0.6745\pm 0.0148$.

Thus we have found that the AT line passes through the point
$(T,h_r)=(0.674,0.1)$. To compare that with the predictions from the
SK model, we use the zero-field transition temperature $T_{c}= 0.887$
obtained above.  Then for $h_r=0.1$, the predicted value of the AT
transition temperature ratio of the SK model (from
Eq.~\ref{eq:approx_AT_line}) would be
$T_{\text{AT}}(h_r=0.1)/T_c = 0.74$, while the Monte Carlo determined
value at $\sigma = 0.6$ is $0.7590\pm 0.0113$. (For the SK model, the
Monte Carlo value of the ratio is $0.7641 \pm 0.0341$). Thus while the
zero-field transition temperature at $\sigma = 0.6$ is not close to
the mean-field value of Eq.~(\ref{eq:T_c^MF_zero_field}), the SK form
of the AT line is a good approximation provided it is expressed in
terms of the renormalized zero-field transition temperature $T_c$ (see
also Fig.~\ref{fig:XY_AT}).

The AT line can be approached not only by reducing the temperature $T$
but also by reducing the field at fixed $T$. In
Figs. \ref{fig:chi_cl_vs_hr_s0.60}\flc{a} and
\ref{fig:chi_cl_vs_hr_s0.60}\flc{b} we have constructed the crossing
plots at fixed temperature $T=0.6$ $(=0.67\,T_c)$ as a function of
$h_r$ for $\chi_{\text{SG}}$ and $\xi_{\text{SG}}$
respectively. Analysis of the crossing plots of $h^*(N, 2N)$ in
Fig. \ref{fig:chi_cl_vs_hr_s0.60}\flc{c} shows that the behavior is
again consistent with the existence of an AT line at least at
$\sigma = 0.60$. The same value of $\lambda$ was used as when plotting
$T^*(N, 2N)$. The $h^*(N,2N)$ data for all the pairs of system sizes
are fitted against $N^{-\lambda}$ to give
$h_{\text{AT}}(T=0.6) = 0.1569 \pm 0.0061$ from $\chi_{\text{SG}}$ and
$h_{\text{AT}}(T=0.6) = 0.1571 \pm 0.0067$ from $\xi_{\text{SG}}$. We
found two points on the AT line, $(T,h_r)=(0.674,0.1)$ from
$T^*(N,2N)$, and $(T,h_r)=(0.6,0.157)$ from $h^*(N,2N)$ data. These
points are plotted in Fig.~\ref{fig:XY_AT} for comparison with the
exact AT line for the SK model.

\subsection{$\sigma=0.75$}
\label{sec:sigma_0.75}

\begin{figure*}
  \centering
  \includegraphics[width=\textwidth]{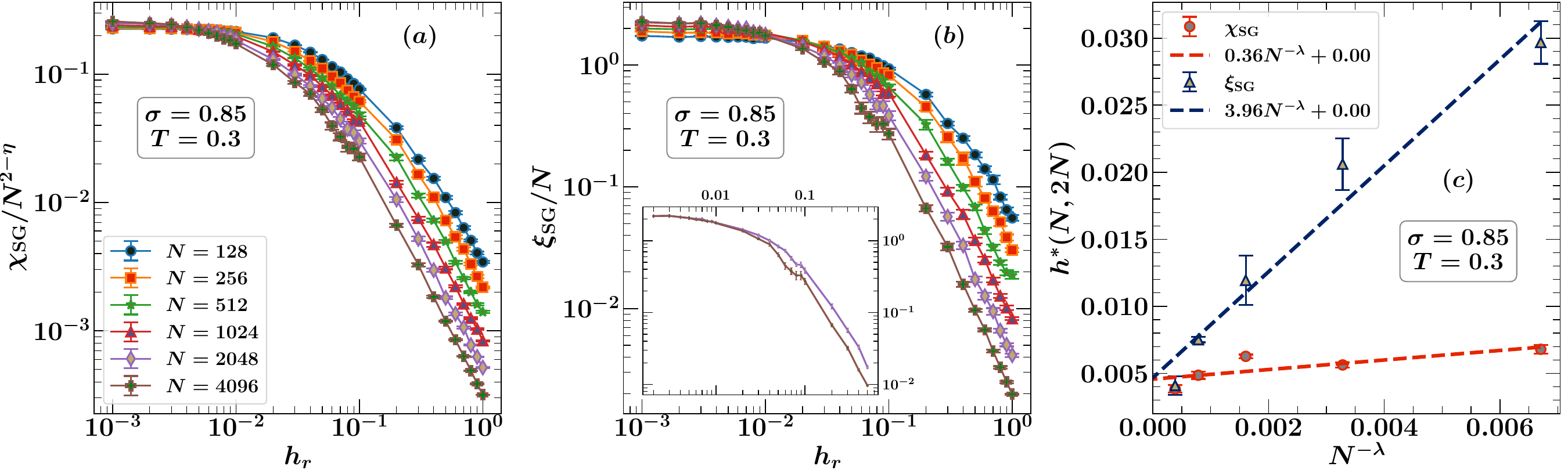}
  \caption{Finite size scaling plots of (a) $\chi_{\text{SG}}$ and (b)
    $\xi_{\text{SG}}$ as a function of magnetic field $h_r$, for
    $\sigma=0.85$ at a temperature of $T=0.3$ $(=0.91\,T_c)$. The
    inset of figure (b) shows the data for the two largest system
    sizes $N=2048$ and $N=4096$. The common legend for both the
    figures (a) and (b) is shown in figure (a). The intersection
    fields $h^*(N,2N)$ data is shown in figure (c). We substitue the
    value of the exponent $\lambda=1.03$ (see Table~\ref{tab:lambda})
    obtained from the $T^*(N,2N)$ data at $h_r=0$ in
    Eq.~(\ref{eq:intersection_field_corrections}) and fit $h^*(N,2N)$
    data agianst $N^{-\lambda}$ to get
    $h_{\text{AT}}(T=0.3) = 0.0046 \pm 0.0006$ from $\chi_{\text{SG}}$
    and $h_{\text{AT}}(T=0.3) = 0.0047 \pm 0.0016$ from
    $\xi_{\text{SG}}$.}
  \label{fig:chi_cl_vs_hr_s0.85}
\end{figure*}

The case $\sigma=0.75$ corresponds to the non-mean-field regime: the
long-range diluted model for this value of $\sigma$ is equivalent to a
short-range model with $d\approx 4$ dimensions.  In this regime,
simulations of the corresponding Heisenberg
model~\cite{sharma2011almeida,sharma2011phase} were thought consistent
with an AT transition.

According to Eq.~(\ref{eq:chi_fss:2/3<sigma<1}), the data for
$\chi_{\text{SG}}/N^{2-\eta}$,where $2-\eta=2\sigma-1$, plotted for
different system sizes should intersect at the transition temperature
$T_c$.  Similarly, according to Eq. (\ref{eq:xi_fss:2/3<sigma<1}), the
curves of $\xi_{\text{SG}}/N$ should also intersect at the transition
temperature. Figs. \ref{fig:chi_cl_vs_T_s0.75}\flc{a} and
\ref{fig:chi_cl_vs_T_s0.75}\flc{d} show the finite-size-scaled data
of $\chi_{\text{SG}}$, and Figs. \ref{fig:chi_cl_vs_T_s0.75}\flc{b}
and \ref{fig:chi_cl_vs_T_s0.75}\flc{e} show the finite-size-scaled
data of $\xi_{\text{SG}}$.  The curves for different system sizes show
a clear tendency to intersect close to the same temperature.  The data
for $T^*(N,2N)$ are then fitted with
Eq. (\ref{eq:intersection_temperature_corrections}) where the value of
the exponent $\lambda$ is not known in the non-mean-field regime and
hence should be considered as a fitting parameter.

If there were an AT transition there would be a unique value of
$\lambda$, the same for both the $\xi_{\text{SG}}$ and
$\chi_{\text{SG}}$ intersections corresponding to both $h^*(N, 2N)$
and $T^*(N, 2N)$. In order to find the value of $\lambda$ through
non-linear fitting, we have six different sets of data: $T^*(N,2N)$
obtained from $\chi_{\text{SG}}$ and $\xi_{\text{SG}}$ intersections,
with $h_r=0$ and $h_r=0.05$ (Figs. \ref{fig:chi_cl_vs_T_s0.75}\flc{a},
\ref{fig:chi_cl_vs_T_s0.75}\flc{b},
\ref{fig:chi_cl_vs_T_s0.75}\flc{d}, and
\ref{fig:chi_cl_vs_T_s0.75}\flc{e}), and $h^*(N,2N)$ obtained from
$\chi_{\text{SG}}$ and $\xi_{\text{SG}}$ intersections at $T=0.55$
(Figs. \ref{fig:chi_cl_vs_hr_s0.75}\flc{a} and
\ref{fig:chi_cl_vs_hr_s0.75}\flc{b}). We tried fitting these
individual data sets with
Eq. (\ref{eq:intersection_temperature_corrections})
(Eq. (\ref{eq:intersection_field_corrections}) for $h^*(N,2N)$)
through non-linear fitting by considering $\lambda$, $T_c$ and $A$
($\tilde{A}$ for $h^*(N,2N)$) as fitting parameters. This is a
non-linear fitting procedure for which we use efficient methods like
the Trusted Region Reflective (TRF) algorithm and the
Levenberg-Marquardt (LM) algorithm (for which packages are available
in python) to determine the fitting parameters. Doing so, we found
that the $h^*(N,2N)$ data obtained from the $\chi_{\text{SG}}$
intersections at $T=0.55$ (Fig. \ref{fig:chi_cl_vs_hr_s0.75}\flc{a})
gave us the best fit (using chi-square test), and we obtain
$\lambda=0.26$ (see Table~\ref{tab:lambda}).
%
%
Since the exponent giving the leading correction to scaling $\lambda$
is universal, we use the same value of $\lambda$ with both
intersections $h^*(N,2N)$ and $T^*(N,2N)$ obtained from
$\chi_{\text{SG}}$ and $\xi_{\text{SG}}$ data. We substitute the value
of $\lambda$ obtained above in
Eq.~(\ref{eq:intersection_temperature_corrections}) and fit the
$T^*(N,2N)$ data against $N^{-\lambda}$ with a straight line. As shown
in Fig.~\ref{fig:chi_cl_vs_T_s0.75}\flc{c}, for $h_r=0$, the
$\chi_{\text{SG}}$ fit (considering all the pairs of system sizes)
gives $T_{c} = 0.6200 \pm 0.0031$. The corresponding $\xi_{\text{SG}}$
fit (omitting the smallest system size) gives
$T_{c} = 0.6244 \pm 0.0098$.

For $h_r=0.05$, the intersection temperatures data are shown in
Fig.~\ref{fig:chi_cl_vs_T_s0.75}\flc{f}. Omitting the smallest system
size, the $T^*(N,2N)$ data are fitted with
Eq. (\ref{eq:intersection_temperature_corrections}) to give
$T_{\text{AT}}(h_r=0.05) = 0.4395 \pm 0.0241$ from $\chi_{\text{SG}}$
and $T_{\text{AT}}(h_r=0.05) = 0.2893 \,\pm\, 0.0252$ from
$\xi_{\text{SG}}$. Compared to Fig. \ref{fig:chi_cl_vs_T_s0.60}\flc{f}
which gives the equivalent plot for the case with $\sigma =0.60$, the
data in Fig. \ref{fig:chi_cl_vs_T_s0.75}\flc{f} does not look like
data which is converging to the same asymptotic limit when $N$ is
large. If the crossings were actually due to a genuine AT transition,
then the asymptotic limit should be the same for both. If we follow
the mean-field prescription (Eq.~\ref{eq:approx_AT_line}) that is
applicable to the SK model, the spin glass transition temperature for
$h_r=0.05$ is $T_{\text{AT}}(h_r=0.05)/T_c=0.83$. The minimum
temperature we simulated for $\sigma=0.75$ is
$T_{\text{min}}/T_c=0.45$ (look at Table~\ref{tab:parameters_T}),
which is $0.54$ times the mean-field prediction. On the other hand,
for $\sigma=0.60$ with $h_r=0.1$, the mean-field calculations give
$T_{\text{AT}}(h_r=0.1)/T_c=0.74$, and the minimum temperature
simulated for this case is $T_{\text{min}}/T_c=0.56$. So, for
$\sigma=0.60$ the minimum temperature simulated is just $76\%$ of the
mean-field transition temperature at that particular field and still
we were able to observe clear signs of a phase transition. In
contrast, for $\sigma=0.75$, we went down to a much lower temperature
which is $54\%$ of the mean-field spin glass transition temperature at
$h_r=0.05$ and still we couldn't see clear signs of a phase
transition.

We have also studied $\chi_{\text{SG}}$ and $\xi_{\text{SG}}$ at fixed
$T$, but varying $h_r$ and the finite size scaling plots for these are
given in Figs. \ref{fig:chi_cl_vs_hr_s0.75}\flc{a} and
\ref{fig:chi_cl_vs_hr_s0.75}\flc{b}. There appears to be good
intersections in the curves, supporting therefore the possible
existence of an AT transition at the temperature studied $T =0.55$
$(=0.89\,T_c)$. A plot of $h^*(N, 2N)$ versus $1/N^{\lambda}$ is in
Fig. \ref{fig:chi_cl_vs_hr_s0.75}\flc{c}, using the same value of
$\lambda=0.26$. In the intersections of $\xi_{\text{SG}}$ there is a
clear rising trend of $h^*(N, 2N)$ with increasing $N$ until
$N = 1024$, followed by decreasing values of $h^*(N, 2N)$ for
$N> 2048$. For the case of $\sigma= 0.60$, where there is almost
certainly a genuine AT transition,
(Fig. \ref{fig:chi_cl_vs_hr_s0.60}\flc{c}) only the rising trend is
seen. It is as if for the smaller systems $N < 2048$ the system at
$\sigma = 0.75$ is behaving similarly to its mean-field cousin at
$\sigma = 0.60$. Note that this change of trend cannot be attributed
to the correction to scaling terms of
Eq.~(\ref{eq:scalingcorrform1}). These only apply in the limit
$N \to \infty$ with $N^{1/\nu}(h_r-h_{\text{AT}}(T))$ fixed.  For a
genuine AT transition the intersections $h^*(N, 2N)$ from both
$\xi_{\text{SG}}$ and $\chi_{\text{SG}}$ should both extrapolate as
$N \to \infty$ to the \textit{same} field $h_{\text{AT}}(T)$. It is
hard to argue that Fig. \ref{fig:chi_cl_vs_hr_s0.75}\flc{c} provides
good evidence for this. On the other hand, on the droplet picture, it
would be expected that $h^*(N, 2N)$ should extrapolate to zero. The
evidence that is happening is also weak.

\subsection{$\sigma=0.85$}
\label{sec:sigma_0.85}

\begin{figure*}
  \centering
  \includegraphics[width=\textwidth]{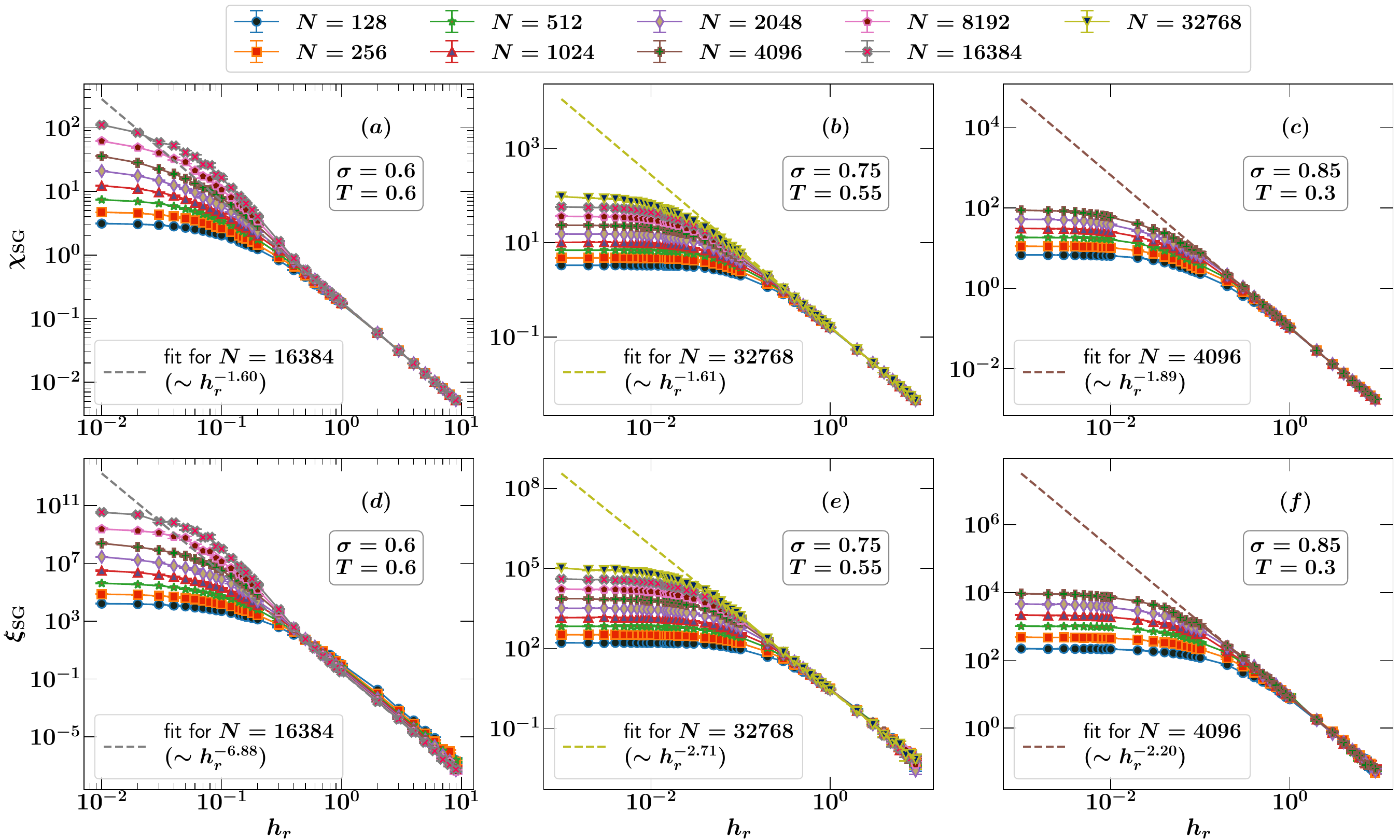}
  \caption{Figures in the top row show $\chi_{\text{SG}}$ plotted as a
    function of magnetic field $h_r$ at fixed temperature, for: (a)
    $\sigma=0.6$ at $T=0.6$, (b) $\sigma=0.75$ at $T=0.55$, and (c)
    $\sigma=0.85$ at $T=0.3$. Figures in the bottom row show plots of
    $\xi_{\text{SG}}$ versus $h_r$ for: (d) $\sigma=0.6$ at $T=0.6$,
    (e) $\sigma=0.75$ at $T=0.55$, and (f) $\sigma=0.85$ at
    $T=0.3$. For large values of $h_r$, we fitted the
    $\chi_{\text{SG}}$ data with $1/h_r^{x_{\chi}}$ and
    $\xi_{\text{SG}}$ data with $1/h_r^{x}$. The color labelling for
    different system sizes is same for all these figures, and the
    common legend is displayed at the top. The values of exponents
    $x_{\chi}$, $x$, and $z=x_{\chi}/x$ obtained from curve fitting
    are presented in Table~\ref{tab:scaling_exponents}.}
  \label{fig:chi_cl_hr_fit}
\end{figure*}

\begin{table}
  \centering  
  \caption{Values of the exponent $\lambda$ obtained from our
    simulations for different values of $\sigma$. The last column
    shows the values of $\chi^2/N_{\text{dof}}$ which is a measure of
    goodness-of-fit. $N_{\text{dof}}$ denotes the number of degrees of
    freedom. For $\sigma=0.75$, the value of $\lambda$ shown here is
    obtained by fitting the $h^*(N,2N)$ data obtained from
    $\chi_{\text{SG}}$ intersections in
    Fig.~\ref{fig:chi_cl_vs_hr_s0.75}\flc{a} with
    Eq.~(\ref{eq:intersection_field_corrections}) using efficient
    non-linear fitting algorithms. For $\sigma=0.85$, we use the
    $T^*(N,2N)$ data from $\chi_{\text{SG}}$ intersections in
    Fig.~\ref{fig:chi_cl_vs_T_s0.85}\flc{a} to determine
    $\lambda$. For convenience, we include a column showing our
    estimates for the zero-field spin glass transition temperature for
    different $\sigma$. These values are obtained from the fits shown
    in Figs. \ref{fig:chi_cl_vs_T_s0.60}\flc{c},
    \ref{fig:chi_cl_vs_T_s0.75}\flc{c}, and
    \ref{fig:chi_cl_vs_T_s0.85}\flc{c}.  \vspace{0.2cm}}
  \label{tab:lambda}
  \begin{tabular*}{0.35\textwidth}{c @{\extracolsep{\fill}} ccc}
    \hline
    \hline 
    $\sigma$ &  $T_c$  & $\lambda$ & $\chi^2/N_{\text{dof}}$ \Tstrut\\[0.15cm]
    \hline
    $0.60$  &  $0.89$  &  $0.47$  &  - \\
    $0.75$  &  $0.62$  &  $0.26$  &  $1.03$ \\
    $0.85$  &  $0.33$  &  $1.03$  &  $0.39$ \\
    \hline
    \hline
  \end{tabular*}    
\end{table}

For $\sigma=0.85$ we are further into the non-mean-field region.
According to Eq.~(\ref{eq:d_eff}), $\sigma=0.85$ corresponds to a
short-range model close to three dimensions.  In this regime,
simulations of the corresponding Heisenberg
model~\cite{sharma2011almeida,sharma2011phase} did not find an AT
line.

For $h_r=0$, Figs. \ref{fig:chi_cl_vs_T_s0.85}\flc{a} and
\ref{fig:chi_cl_vs_T_s0.85}\flc{b} clearly show that the curves for
different system sizes are intersecting.  The data for intersection
temperatures are shown in Fig. \ref{fig:chi_cl_vs_T_s0.85}\flc{c}.
Similar to the case of $\sigma=0.75$, the $T^*(N,2N)$ data obtained
from $\chi_{\text{SG}}$ and $\xi_{\text{SG}}$ intersections with
$h_r=0$ (Figs. \ref{fig:chi_cl_vs_T_s0.85}\flc{a} and
\ref{fig:chi_cl_vs_T_s0.85}\flc{b}), and the $h^*(N,2N)$ obtained from
$\chi_{\text{SG}}$ and $\xi_{\text{SG}}$ intersections at $T=0.3$
(Figs. \ref{fig:chi_cl_vs_hr_s0.85}\flc{a} and
\ref{fig:chi_cl_vs_hr_s0.85}\flc{b}) are fitted with
Eq. (\ref{eq:intersection_temperature_corrections})
(Eq. (\ref{eq:intersection_field_corrections}) for $h^*(N,2N)$) by
considering $\lambda$, $T_c$, and $A$ ($\tilde{A}$ for $h^*(N,2N)$) as
fitting parameters, and found that the $T^*(N,2N)$ data obtained from
the $\chi_{\text{SG}}$ intersections gave us the best fit. We obtain
$\lambda=1.03$ (see Table~\ref{tab:lambda}) from both TRF and LM
methods. We use this value of $\lambda$ in
Eqs. (\ref{eq:intersection_temperature_corrections}) and
(\ref{eq:intersection_field_corrections}) for further
calculations. The fit using the $\chi_{\text{SG}}$ data for all the
pairs of system sizes gives $T_{c} = 0.3336 \pm 0.0013$. The
corresponding $\xi_{\text{SG}}$ fit (omitting the smallest system
size) gives $T_{c} = 0.3297 \pm 0.0036$.  The two values of $T_{c}$
are quite close.

For $h_r=0.05$ the $\chi_{\text{SG}}/N^{2-\eta}$ data do not intersect
as shown in Fig. \ref{fig:chi_cl_vs_T_s0.85_infield}\flc{a}. Such a
field could conceivably be above the largest AT field even at $T= 0$,
so we also studied a smaller field: $h_r=0.02$ shown in
Fig. \ref{fig:chi_cl_vs_T_s0.85_infield}\flc{c}. There is no sign of
any crossing at this field either!.  The $\xi_{\text{SG}}$ data is
less clearcut. Fig. \ref{fig:chi_cl_vs_T_s0.85_infield}\flc{b} shows
there are no intersections at a field of $h_r=0.05$ while a merging
behavior is seen for the larger systems at $h_r=0.02$, as shown in
Fig. \ref{fig:chi_cl_vs_T_s0.85_infield}\flc{d}. In our simulations we
went to very low temperatures such as $T=0.1$, which is small in
comparison with the mean-field values of $T_{\text{AT}}$ for
$h_r=0.02$ and $h_r=0.05$ using Eq. (\ref{eq:approx_AT_line}), but we
still could not find any clear intersections in the $\chi_{\text{SG}}$
or $\xi_{\text{SG}}$ data.  This suggests that there is no phase
transition in this regime in the presence of a magnetic field. Our
data are consistent with the scenario where the external magnetic
field destroys the phase transition, just as happens for a ferromagnet
when a uniform field is turned on. Very similar features were seen for
the Heisenberg version of this model
\cite{sharma2011almeida,sharma2011phase} and in the three dimensional
Ising model \cite{Baity_Jesi_2014}.

Confusingly, intersections are seen at fixed $T= 0.3$ $(=0.91\,T_c)$
as $h_r$ is varied in the plots of $\chi_{\text{SG}}/N^{2-\eta}$ in
Fig. \ref{fig:chi_cl_vs_hr_s0.85}\flc{a} and of $\xi_{\text{SG}}/N$ in
Fig. \ref{fig:chi_cl_vs_hr_s0.85}\flc{b}. The usual analysis of
$h^*(N, 2N)$ is given in
Fig. \ref{fig:chi_cl_vs_hr_s0.85}\flc{c}. Thus in crossing the AT line
along a trajectory of fixed $T$ we have seen intersections, suggesting
there might be an AT transition.  However, the large $N$ limit of
$h^*(N, 2N)$ in Fig. \ref{fig:chi_cl_vs_hr_s0.85}\flc{c} in the case
of $\sigma = 0.85$, suggests that $h_{\text{AT}}(T)$ might actually be
zero, consistent with the droplet scaling picture. In the next section
the dependence of $\xi_{\text{SG}}$ and $\chi_{\text{SG}}$ on $h_r$
will be explained using the droplet scaling approach.

\section{Data analyses on the droplet picture}
\label{sec:dropletforms}

\begin{figure*}
  \centering
  \includegraphics[width=\textwidth]{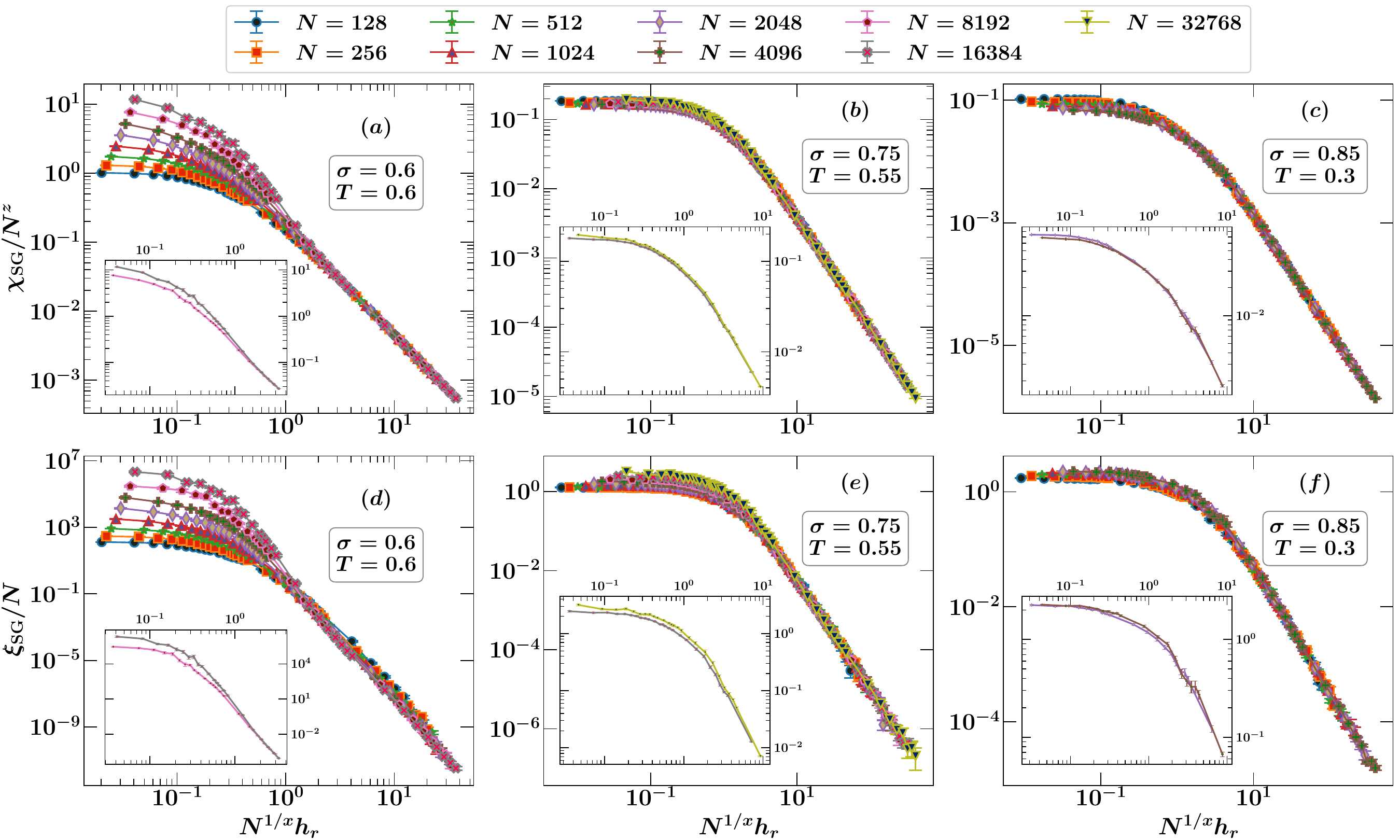}
  \caption{Finite size scaling analysis of $\chi_{\text{SG}}$ data
    (figures (a),(b),(c)) and $\xi_{\text{SG}}$ data (figures
    (d),(e),(f)) on the droplet picture, obtained by fixing the
    temperature $T$ and varying the magnetic field $h_r$. The data are
    plotted on a log-log scale.  Figures (a) and (d) are for
    $\sigma=0.6$ at $T=0.6$, (b) and (e) are for $\sigma=0.75$ at
    $T=0.55$, and (c) and (f) are for $\sigma=0.85$ at $T=0.3$. The
    corresponding inset figures show our data for the two largest
    system sizes: $N=8192 \text{ and } 16384$ for $\sigma=0.6$,
    $N=16384 \text{ and } 32768$ for $\sigma=0.75$, and
    $N=2048 \text{ and } 4096$ for $\sigma=0.85$. The legend displayed
    at the top is common for all the figures (both main and
    inset). The values of the exponents $x$ and $z$ are given in
    Table~\ref{tab:scaling_exponents}.}
  \label{fig:chi_cl_Nxhr}
\end{figure*}

In this section we give the field dependence of $\xi_{\text{SG}}$ and
$\chi_{\text{SG}}$ according to the droplet picture
\cite{McMillan_1984, bray:86, FisherHuse:88}, including also their
finite size modifications, and compare these with our simulation data.

In the droplet picture one uses an Imry-Ma argument
\cite{PhysRevLett.35.1399} for the correlation length $\xi$ and
identifies it with the size of the region or domain within which the
spins become re-oriented in the presence of the random field. The free
energy gained from such a reorientation by the the random field is of
order $\sqrt{q_{\text{EA}}(T)}h_r \xi^{d/2}$. The size of such domains
$\xi$ is determined by equating this free energy to the free energy
cost of the interface of this domain of re-ordered spins with the rest
of the system, which is of the form $\Upsilon(T) \xi^{\theta}$ ~\cite
{PhysRevE.93.032123}. Equating these two free energies gives
\begin{equation}
\label{eq:Imry-Malength}
\xi \sim \bigg[\frac{\Upsilon(T)}{\sqrt{q_{\text{EA}}(T)}h_r}\bigg]^{1/(d/2-\theta)}.
\end{equation}
While there is a considerable literature on the dependence of the
interface exponent $\theta$ on $\sigma$ for the case of Ising spin
glasses ~\cite{moore:16}, we know of no equivalent studies for the
case of the XY spin glass. (Our data suggests that its $\theta$ might
be close to that of the Ising spin glass).

Eq.~(\ref{eq:Imry-Malength}) shows that as $h_r\to 0$, the length
scale becomes infinite; $\xi$ diverges as $\xi \sim 1/h_r^x$, where
\begin{equation}
x = \frac{1}{d/2-\theta}.
\label{eq:xdef}
\end{equation}
The exponent $x$ is the analogue of $\nu$ at the AT transition; it is
as if the AT transition $h_{\text{AT}}(T)=0$.  We would expect this
formula to apply until finite size effects limit its growth, which
will occur when $\xi$ is of $O(L)$ (or $O(N)$ in our one-dimensional
system). Identifying $\xi_{\text{SG}}$ with $\xi$,
Figs. \ref{fig:chi_cl_hr_fit}\flc{e} and
\ref{fig:chi_cl_hr_fit}\flc{f} show that the Imry-Ma fit indeed works
well at the larger fields; the data for the larger $h_r$ collapse
nicely onto a power law form as predicted by
Eq. (\ref{eq:Imry-Malength}) for all sizes $N$. It only departs from
this formula when $\xi_{\text{SG}}$ becomes of order $N$, when finite
size corrections to the Imry-Ma formula are needed. Also TNT effects
(see Sec.~\ref{TNT}) produce corrections to the Imry-Ma formula when
$\xi_{\text{SG}}$ is of $O(N)$ unless $N= L > L^*$.  The crossover
scale $L^*$ is thought to be large, especially as $\sigma$ approaches
$2/3$ (or $d \to 6$) \cite{Moore:21}.

\begin{figure}
  \centering
  \includegraphics[width=0.48\textwidth]{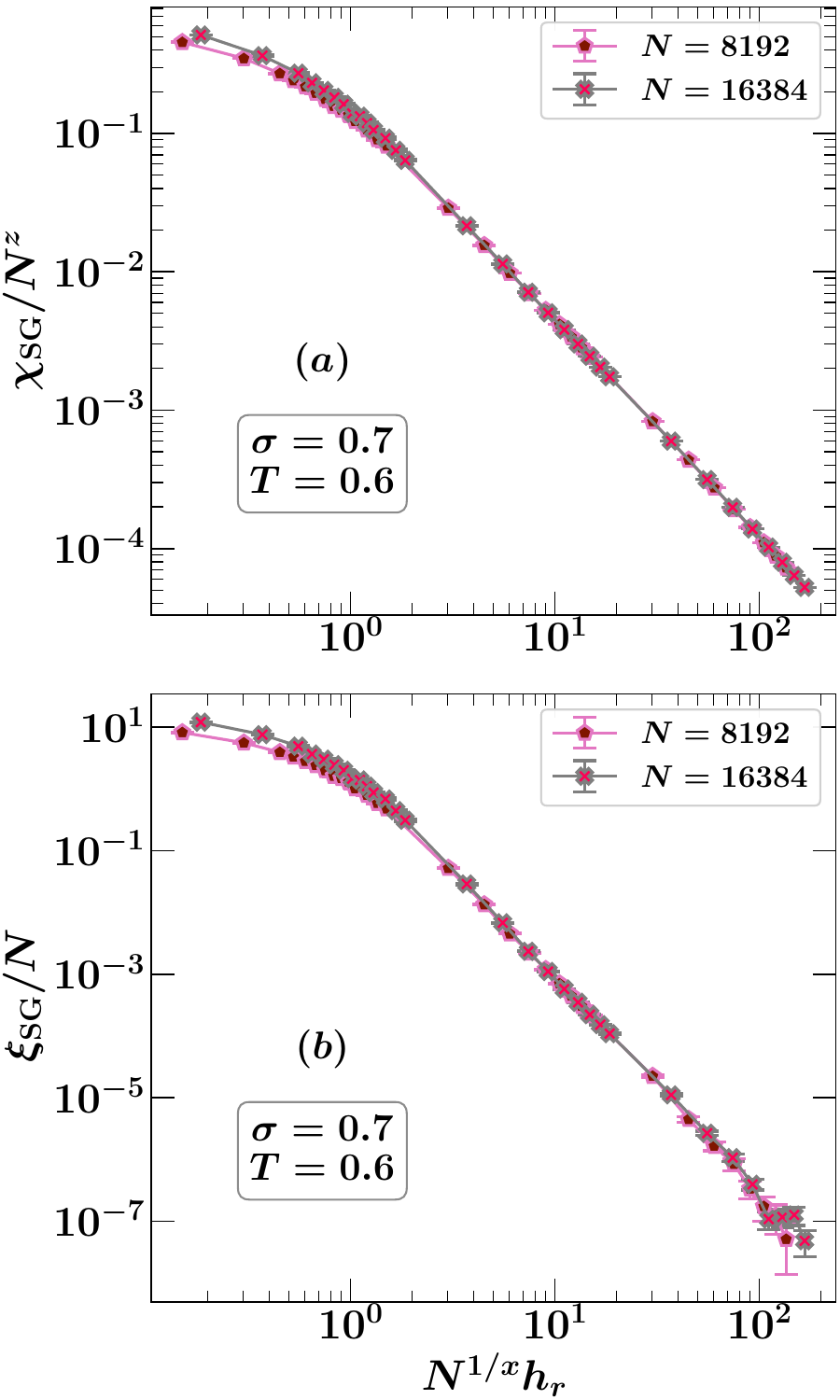}
  \caption{A complete finite size scaling plot of (a)
    $\chi_{\text{SG}}$ and (b) $\xi_{\text{SG}}$ as a function of
    magnetic field $h_r$, for $\sigma=0.70$ at a temperature of
    $T=0.6$ $(=0.83\,T_c,\,T_c=0.72)$ for our two largest system
    sizes.}
  \label{fig:chi_cl_Nxhr_s0.70}
\end{figure}

To allow for finite size effects on the Imry-Ma formula we use the
analogue of Eq.~(\ref{eq:chi_fss:2/3<sigma<1h}) with $h_{\text{AT}}=0$ and
$\nu = x$ to write:
\begin{equation}
\frac{\xi_{\text{SG}}}{N}=\mathcal{X}(N^{1/x} h_r).
\label{eq:xi_fss_IM}
\end{equation}
Our results for $\sigma = 0.75$ are shown in
Fig.~\ref{fig:chi_cl_Nxhr}\flc{e} and for $\sigma=0.85$ are shown in
Fig.~ \ref{fig:chi_cl_Nxhr}\flc{f}. There are clearly finite size
corrections to this formula. It is a formula which formally would be
expected to hold in the scaling limit of $N \to \infty$ with
$N^{1/x} h_r$ fixed. The crossover function
$\mathcal{X}(y) \sim 1/y^{x}$ when $y$ is large, in order to recover
Eq.~(\ref{eq:Imry-Malength}). It goes to a constant when $y \to
0$. However, a closer look at our two largest system sizes $N = 16384$
and $N= 32768$ at $\sigma = 0.75$ (inset to
Fig.~\ref{fig:chi_cl_Nxhr}\flc{e}) and our two largest system sizes at
$\sigma = 0.85$, $N= 2048$ and $N= 4096$ (inset to
Fig.~\ref{fig:chi_cl_Nxhr}\flc{f}) shows that the finite size
corrections are becoming small, and are smaller the further the system
is away from the mean-field region. If one moves in the other
direction, towards the start of the mean-field region $\sigma = 2/3$,
the finite size corrections are larger, as seen in
Fig. \ref{fig:chi_cl_Nxhr_s0.70}\flc{b} for $\sigma = 0.70$. The finite
size scaling form for these corrections to the scaling of
Eq.~(\ref{eq:xi_fss_IM}) will be of the form
\begin{equation}
\frac{\xi_{\text{SG}}}{N}=\mathcal{X}( N^{1/x}h_r)+ N^{-\omega}\mathcal{H}( N^{1/x}h_r),
\label{eq:scalingcorrformxi}
\end{equation}
where $\omega$ is the correction to scaling exponent. However, TNT
effects (see Sec.~\ref{TNT}) produce large further corrections to
these asymptotic forms when $L < L^*$. Since in our studies $L^*$ is
probably larger than the length $N$ of our system, at least for
$\sigma = 0.75$, the scaling form of Eq.~(\ref{eq:scalingcorrformxi})
does not work in the region where $\xi_{\text{SG}}$ is of order $N$
(see Fig.~\ref{fig:chi_cl_Nxhr_linear}\flc{a}). For $\sigma =0.85$
where $L^*$ is expected to be smaller,
Fig.~\ref{fig:chi_cl_Nxhr_linear}\flc{b} hints that
Eq.~(\ref{eq:scalingcorrformxi}) might apply as the plots at adjacent
sizes for the larger $N$ values seem to be getting closer together as
$N$ is increased, which is a feature predicted by
Eq.~(\ref{eq:scalingcorrformxi}).

In Fig. \ref{fig:chi_cl_hr_fit}\flc{d} we show a similar plot to those
in Figs. \ref{fig:chi_cl_hr_fit}\flc{e} and
\ref{fig:chi_cl_hr_fit}\flc{f} but for the case of $\sigma =
0.60$. Notice however that because of the AT transition at this value
of $\sigma$, at which $\xi_{\text{SG}}$ would diverge to infinity as
$N\to \infty$ at some finite field $h_r =h_{\text{AT}}(T)$, a shoulder
above the dashed line has started to appear which is the beginning of
this divergence. Such a feature is absent in the figures for both
$\sigma = 0.75$ and at $\sigma = 0.85$. Similarly,
Fig. \ref{fig:chi_cl_Nxhr}\flc{d} shows poor collapse of
$\xi_{\text{SG}}$ data for $\sigma=0.6$, which is in contrast to the
cases of $\sigma=0.75$ and $\sigma=0.85$. This indicates that the
$\xi_{\text{SG}}$ data for $\sigma=0.6$ is not in accordance with
Eq. \ref{eq:scalingcorrformxi}.

\begin{figure}
  \centering
  \includegraphics[width=0.48\textwidth]{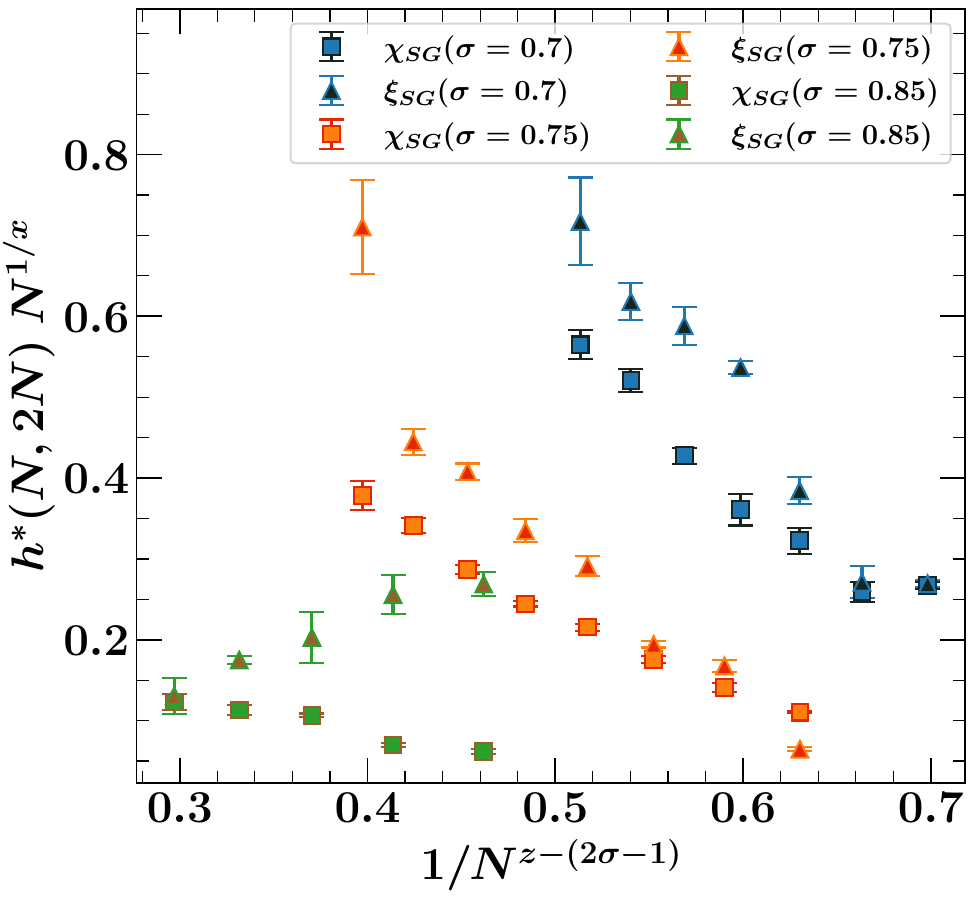}
  \caption{A plot of $h^*(N, 2N)N^{1/x}$ versus $1/N^{\omega}$, for
    $\sigma = 0.70, 0.75$ and $0.85$. The values of $x$ and $\omega$,
    which is obtained from Eq.~(\ref{eq:omegaspec}) were taken from
    Table~\ref{tab:scaling_exponents}.}
  \label{fig:ifNx}
\end{figure}

The spin glass susceptibility according to the droplet picture is a
similar generalization of the finite size scaling form of
Eq.~(\ref{eq:chi_fss:2/3<sigma<1}):
\begin{equation}
\frac{\chi_{\text{SG}}}{N^z} = \mathcal{C}( N^{1/x}h_r),\quad
                                          (2/3 \leq \sigma < 1). 
\label{eq:chiSGIM}
\end{equation}
The crossover function $\mathcal{C}(y) \sim 1/y^{x_{\chi}}$ when $y$ is large, so that
then $\chi_{\text{SG}} \sim \xi^z$ and becomes independent of $N$.
Its form is then
\begin{equation}
\chi_{\text{SG}}\sim 1/h_r^{x_{\chi}},
\label{eq:xchidef}
\end{equation}
which implies that $x_{\chi}= xz$.  In the opposite limit as
$y \to 0$, $C(y)$ goes to a finite constant.
Figs. \ref{fig:chi_cl_hr_fit}\flc{a}, \ref{fig:chi_cl_hr_fit}\flc{b},
and \ref{fig:chi_cl_hr_fit}\flc{c} show that the $\chi_{\text{SG}}$
data for the larger $h_r$ collapse nicely onto a power law form as
predicted by Eq. (\ref{eq:xchidef}) for all sizes $N$. The exponent
$z$ depends upon whether we are dealing with short-range interactions,
(such as nearest-neighbor interactions) or with the long-range
interactions employed in this paper. For short-range interactions, the
average value of $\chi_{ij}^2$ falls off with spin separation $r_{ij}$
as
\begin{equation}
  \overline{\chi_{ij}^2} \sim \frac{\left(q_{\text{EA}}(T)\right)^2 T}{\Upsilon(T) r_{ij}^\theta},
\label{eq:replicon}
\end{equation}
 \cite{bray:86,FisherHuse:88}. This result applies in the zero-field spin glass state. Then as, 
\begin{equation}
\chi_{\text{SG}}= \frac{1}{N}\sum_{i,j=1}^N \overline{\chi_{ij}^2},
\label{eq:chiSGform}
\end{equation} 
so in $d$ dimensions for the zero-field spin glass $\chi_{\text{SG}} \sim L^{d-\theta}$. Hence
\begin{equation}
z =\frac{d-\theta}{d},
\label{eq:zSR}
\end{equation}
in order to recover the result $\chi_{\text{SG}} \to N^z$ as $N^{1/x} h_r$ goes to zero. We caution that this formula for $z$ will only hold for short-range interactions.
  
With long-range interactions a ``droplet'' is not a single connected
region but a set of isolated islands of flipped spins \cite{moore:16}
and this will make the decay of $\chi_{ij}^2$ with $r_{ij}$ faster
than in Eq.~(\ref{eq:replicon}). This is an effect which has not been
studied before, and so in our problem the exponent $z$ has to be
determined by fitting the data. The results of our determinations of
the droplet exponents $x$, $x_{\chi}$ and $z$ for the different values
of $\sigma$ which we have studied are summarised in
Table~\ref{tab:scaling_exponents}.

The resulting excellent data collapse (at least when
$\xi_{\text{SG}} <N$), is shown in Figs. \ref{fig:chi_cl_Nxhr}\flc{b}
and \ref{fig:chi_cl_Nxhr}\flc{c}. The value of $z$ was determined from
the observation that when $N^{1/x} h_r$ is large, $\chi_{\text{SG}}$
should be independent of $N$. It is remarkable that $z$ determined at
large values of $N^{1/x}h_r$ results in a decent collapse of the data
in the opposite limit where $N^{1/x}h_r \to 0$. Nevertheless
corrections to the Imry-Ma scaling form are visible in the figures
(and are sizeable in the region where $N^{1/x} h_r$ is small when
viewed in a linear plot rather than a log scale plot, (just as in the
$\xi_{\text{SG}}$ plots Figs. \ref{fig:chi_cl_Nxhr}\flc{e} and
\ref{fig:chi_cl_Nxhr}\flc{f}). In the limit when $N^{1/x}h_r$ is held
fixed with $N \to \infty$ the leading correction to scaling will be
\begin{equation}
\frac{\chi_{\text{SG}}}{N^z} = \mathcal{C}( N^{1/x}h_r)+N^{-\omega}\mathcal{G}(N^{1/x} h_r),
\label{eq:scalingcorrformchi}                                        
\end{equation}
where $\mathcal{G}(y)$ is an unknown scaling function and the
correction to scaling exponent $\omega$ is not known with any
certainty (but see Eq.~(\ref{eq:omegaspec})).

\begin{table}[t]
  \centering  
  \caption{Values of the exponents $x_{\chi}$, $x$, and $z$ obtained
    from our simulations for different values of $\sigma$ and $T$
    (look at Fig. \ref{fig:chi_cl_hr_fit}).}
  \label{tab:scaling_exponents}
  \begin{tabular*}{0.5\textwidth}{c @{\extracolsep{\fill}} cccc}
    \hline
    \hline 
    $\sigma$ & $T$ & $x_{\chi}$ & $x$  & $z$ \Tstrut\\[0.15cm]
    \hline
    
    $0.7$   & $0.6$   & $1.5747 \pm 0.0009$ & $3.3220 \pm 0.0413$ & $0.4740 \pm 0.0062$ \\
    $0.75$  & $0.55$  & $1.6114 \pm 0.0005$ & $2.7077 \pm 0.0531$ & $0.5951 \pm 0.0119$ \\
    $0.85$  & $0.3$   & $1.8919 \pm 0.0014$ & $2.2019 \pm 0.0152$ & $0.8592 \pm 0.0066$ \\
    \hline
    \hline
  \end{tabular*}    
\end{table}

\begin{figure*}
  \centering
  \includegraphics[width=\textwidth]{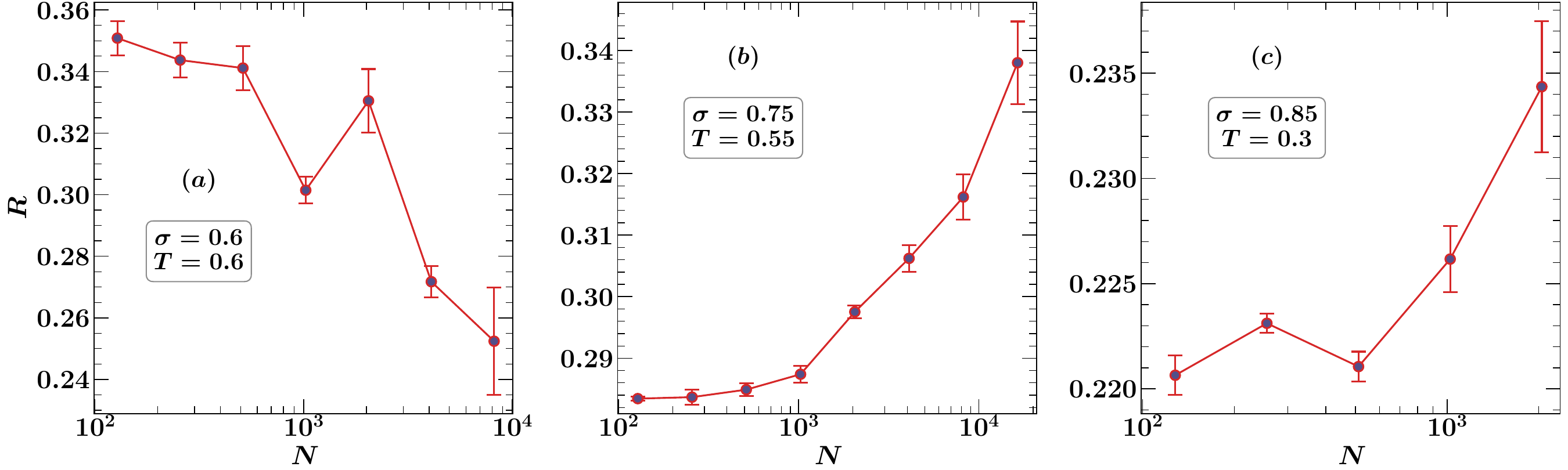}
  \caption{A plot of $R$ (Eq.~(\ref{eq:Rdef})) versus $N$, obtained
    from the data generated by varying the magnetic field $h_r$ at a
    fixed temperature $T$. The data is shown for: (a) $\sigma=0.6$ at
    $T=0.6$, (b) $\sigma=0.75$ at $T=0.55$, and (c) $\sigma=0.85$ at
    $T=0.3$. The quantity $R$ is the y-coordinate of the point where
    the curves for adjacent sizes in
    Figs. \ref{fig:chi_cl_vs_hr_s0.60}\flc{a},
    \ref{fig:chi_cl_vs_hr_s0.75}\flc{a}, and
    \ref{fig:chi_cl_vs_hr_s0.85}\flc{a} intersect.}
  \label{fig:Rxlog}
\end{figure*}

Let us suppose that the droplet picture is correct and that (say) the
spin glass susceptibility $\chi_{\text{SG}}$ is described by
Eq.~(\ref{eq:chiSGIM}). This equation predicts that there will be a
\textit{crossing} in the plots of $\chi_{\text{SG}}/N^{2\sigma-1}$
used in AT line critical scaling studies. (Note we are setting
$2-\eta= 2\sigma-1$). The correction to scaling term of
Eq.~(\ref{eq:scalingcorrformchi}) is not needed for this, but this
correction does strongly influence where the crossings take place for
the $N$ values which are reached in our simulations.  The crossing
arises as follows. At small values of $h_rN^{1/x}$, the function
$\mathcal{C}$ goes to a constant. It turns out that $z > (2\sigma-1)$,
so $\chi_{\text{SG}}/N^{2\sigma-1}$ diverges as $N$ is increased as
$N^{z-(2\sigma-1)}$ as $h_r\to 0$.  On the other hand when
$h_rN^{1/x}$ is large, $\chi_{\text{SG}}\to 1/h_r^z$, so
$\chi_{\text{SG}}/N^{2\sigma-1}\to 1/N^{2\sigma-1}h_r^z \to 0$ as $N$
goes to infinity. Because at small fields,
$\chi_{\text{SG}}/N^{2\sigma-1}$ is larger for large $N$, but at
bigger $h_r$ fields it is smaller at the larger $N$ values, so there must
be a crossing point. We shall denote the crossing value between the
lines at $N$ and $2N$ by $h^*(N, 2N) =H$. Then $H$ is determined by
the solution of the following
\begin{eqnarray}
&&\frac{\chi_{\text{SG}}(H, N)}{N^{2 \sigma-1}}=\mathcal{C}(N^{1/x}H) N^{z-(2\sigma-1)}=\nonumber
\\&&\frac{\chi_{\text{SG}}(H,2N)}{(2 N)^{2 \sigma-1}}=\mathcal{C}((2N)^{1/x} H) (2N)^{z-(2\sigma-1)}.
\label{eq:h*chi}
\end{eqnarray}
Assuming $\mathcal{C}(y) \rightarrow a-b y$, when $y\to 0$, it is easy
to show then that the $N$ dependence of $h^*(N, 2N)$ at very large $N$
will be as $1/N^{1/x}$.  In reality we have no data in this region of
very large $N$ where the corrections to scaling term in
Eq.~(\ref{eq:scalingcorrformchi}) can be ignored.  The corrections to
scaling are numerically small but are very important in determining
the values of $h^*(N, 2N)$.

There is a similar crossing predicted in the plots of
$\xi_{\text{SG}}/N$ as a function of $h_r$ when
Eq.~(\ref{eq:scalingcorrformxi}) holds, using the analogue of
Eq.~(\ref{eq:h*chi}). In this case it is the scaling correction which
causes the curves to cross, (which requires $\mathcal{H}(0)$ to be
negative), and for these curves the crossings $h^*(N, 2N)$ at very
large $N$ will decrease as $1/N^{1/x+\omega}$, (compare with
Eq.~(\ref{eq:lambda_nonmeanfield})) on taking $\chi(y)=c-dy$ and
$\mathcal{H}(y)\to \text{constant}$ as $y\to 0$. Once again we have no
data in this very large $N$ regime. In Fig. \ref{fig:ifNx} we have
plotted $h^*(N, 2N)N^{1/x}$ versus $1/N^{\omega}$, assuming that
$\omega$ is given by Eq.~(\ref{eq:omegaspec}). Note that the size of
the corrections to scaling $\sim 1/N^{\omega}$ is simply not small for
the values of $N$ which we can study, contrary to what was assumed in
the above. $h^*(N, 2N)N^{1/x}$ should go to a constant as $N$ goes to
infinity and it is only for the case of $\sigma = 0.85$, where the
corrections to scaling are the smallest of the three cases studied,
does that look remotely possible. For the case of $\sigma = 0.70$ the
corrections look to be very large.  We conclude that for the values of
$\sigma = 0.70$ and $\sigma = 0.75$, the crossing data on $h^*(N, 2N)$
is not close to the large $N$ asymptotic form predicted by the
droplet picture. But the droplet picture does predict that the
existence of such intersections.

If we only had information on the values of the crossing fields
$h^*(N, 2N)$ it would be difficult to really be sure whether the
droplet picture or the RSB picture best described the data. The
results on $h^*(N, 2N)$ alone are inconclusive as regards both the AT
transition line picture and the droplet picture.  While on the droplet
picture $h^*(N, 2N)$ are predicted to go to zero as $N \to \infty$,
the values of $h^*(N, 2N)$ are not convincingly going to zero as $N$
is increased (see Fig. \ref{fig:ifNx}). Fortunately, there is another
way of distinguishing the two approaches, which does not require us to
reach the $N$ values at which $h^*(N, 2N)$ starts to approach zero. We
define
\begin{equation}
  R \equiv
  \begin{cases}
    \dfrac{\chi_{\text{SG}}(h^*(N, 2N), N)}{N^{2\sigma-1}}, &2/3 \leq \sigma <1,\\[0.4cm]
    \dfrac{\chi_{\text{SG}}(h^*(N, 2N), N)}{N^{1/3}}, &1/2 < \sigma \leq 2/3.
  \end{cases}  
\label{eq:Rdef}
\end{equation}
(Because we only determine $\chi_{\text{SG}}(h_r, N)$ at a finite
number of values of $h_r$, we use linear interpolation to calculate
$\chi_{\text{SG}}(h^*(N,2N),N)$ using the $\chi_{\text{SG}}(h_r, N)$
values at the two determined values of $h_r$ which lie on either side
of $h^*(N, 2N)$).  On the phase transition picture, $R$ should
approach a finite constant as $N \to \infty$. On the droplet picture
$R$ should increase as $N^{z-(2\sigma-1)}$ as $N\to \infty$. For
$\sigma= 0.60$ where an AT line is expected $R$ should go to a
constant but at the $N$ values studied it actually still appears to be
decreasing (see Fig. ~\ref{fig:Rxlog}\flc{a}) and has yet to become
constant, presumably due to finite size effects. This indicates that
trying to determine whether $\sigma = 2/3$ is the exact value at which
the crossover to droplet scaling behavior will also be challenging
from the side below $2/3$.  However, for $\sigma =0.75$,
Fig.~\ref{fig:Rxlog}\flc{b} shows that $R$ is clearly increasing with $N$
for large $N$ values. But if we had had only data for system sizes
$< 2048$ we might have indeed concluded that there was good evidence
for an AT transition in that $R$ seemed to be an $N$ independent
constant. While at the sizes we can reach $R$ is clearly increasing
with $N$ it has yet to reach its asymptotic form of increase as
$N^{z-(2\sigma-1)}$.  The quantity $R$ also increases with $N$ for
$\sigma = 0.70$ and $\sigma = 0.85$, (see for example
Fig.~\ref{fig:Rxlog}\flc{c}).

In order for $\chi_{\text{SG}}$ to match as $\sigma \to 2/3$ from either
the mean-field side, (where $z = 1/3$) with its value in the non-mean
field region, we would expect that $z$ should approach $1/3$ as
$\sigma \to 2/3$ from above. At $\sigma = 0.85$, $z\approx 0.8409$, at
$\sigma =0.75$, $z\approx 0.6065$, while at $\sigma = 0.70$, we find
$z \approx 0.4737$. Thus it seems quite plausible that $z$ could
approach $1/3$ as $\sigma \to 2/3$ from above. Then the combination
$z-(2\sigma-1)$ would approach zero in this limit, which means that
the divergence of $R$ with $N$ will become harder and harder to see as
$\sigma$ approaches $2/3$. We conclude that it will be challenging to
do numerical work which shows that the AT line disappears at precisely
$\sigma =2/3$. On the mean-field side of $2/3$ the correction to
scaling exponent $\omega=1/3-(2\sigma-1)$. It therefore seems natural
to expect that on the non-mean field regime
\begin{equation}
\omega=z-(2\sigma-1).
\label{eq:omegaspec}
\end{equation}
If valid, this would imply that corrections to scaling should be
larger at $\sigma = 0.70$ than at $\sigma =0.75$, and this is what we
observed in Figs. \ref{fig:chi_cl_Nxhr_s0.70}\flc{b} and
\ref{fig:chi_cl_Nxhr_s0.70}\flc{a}, in comparison with (inset of)
Figs. \ref{fig:chi_cl_Nxhr}\flc{e} and \ref{fig:chi_cl_Nxhr}\flc{b}.

In the presence of a genuine AT transition, as $h_r$ is reduced one
would pass through three regions: first the paramagnetic state at
larger values of $h_r$, then the critical region, then the
low-temperature phase with RSB at smaller values of $h_r$.  The good
data collapse for all values of $h_r$ using Eq.~(\ref{eq:xi_fss_IM}),
and Eq.~(\ref{eq:chiSGIM}) shows that at any finite value of $h_r$
there is just one region, the paramagnetic region. Studying
``intersections'' as in Sec. ~\ref{sec:results} is an attempt to find
the critical region. But the intersections at finite values of $h_r$
for $\sigma =0.75$ and $\sigma = 0.85$ are not signs of a genuine
phase transition, but at least in the case of $\chi_{\text{SG}}$ these
crossings are also just a consequence of droplet scaling. The behavior
of $h^*(N, 2N)$ as a function of $N$ is greatly complicated by finite
size effects and will only become clear at much larger $N$ values than
those which we have been able to study.

Because on the droplet picture there is no AT line and so one is
always in the paramagnetic phase at any non-zero field (just as in a
ferromagnet). However, length scales like $\xi_{\text{SG}}$ become
very large as $h_r\to 0$ for temperatures $T < T_c(h_r=0)$. Once they
become comparable to the system dimensions $L$ and one is in the
regime $h_r < h^*(N, 2N)$, the system will have many of the features
which might be associated with being in the broken replica symmetric
phase which is envisaged to exist below the AT line. For physical
systems in three dimensions the relevant length scale is not the
linear dimension of the system $L$, but the linear dimension of a
fully equilibrated region. This may explain why both simulations and
experiments have failed for many years to resolve the debate.

Might it be possible to find by simulations whether the borderline
between RSB ordering and droplet ordering is at $\sigma = 2/3$, which
is the equivalent of $d =6$ with short-range interactions? To this end
we looked at the case of $\sigma = 0.70$. We found from studying the
crossings of $\xi_{\text{SG}}$ and $\chi_{\text{SG}}$ for the zero
field case that the zero field transition temperature is
$\approx 0.724$. Figs. \ref{fig:chi_cl_Nxhr_s0.70}\flc{a} and
\ref{fig:chi_cl_Nxhr_s0.70}\flc{b} show our attempt to collapse the
data with the droplet scaling forms. Clearly the effects of
corrections to scaling are larger than was the case at $\sigma = 0.75$
in Figs. \ref{fig:chi_cl_Nxhr}\flc{e} and
\ref{fig:chi_cl_Nxhr}\flc{b}. This is in accord with
Eq.~(\ref{eq:omegaspec}) which predicts that the correction to scaling
exponent $\omega$ will go to zero as $\sigma \to 2/3$ if also
$z\to 1/3$ as expected. We conclude that it will be difficult to
provide good numerical evidence that $\sigma = 2/3$ is the lower
critical dimension of the AT transition.

\section{TNT versus The Droplet Scaling Picture}
\label{TNT}

Newman and Stein \cite{newman:03} (see also the recent review
\cite{newman2022metastates}),
have suggested that the ordered phase of spin
glasses in finite dimensions will fall into one of 4 categories, (and
which one might depend on the dimensionality $d$ of the system): The
RSB state is one of these, and is somewhat similar to that envisaged
by Parisi for the SK model, but there is also the chaotic pairs state
picture of Newman and Stein. In both of these pictures there is an AT
transition. The other two pictures are the so-called TNT picture of
Krzakala and Martin \cite{krzakala:00} and Palassini and Young
\cite{palassini:00} and the droplet scaling picture
\cite{McMillan_1984, bray:86, FisherHuse:88}. In neither the TNT
picture nor the droplet scaling picture is there an AT transition. In
the droplet picture the Parisi overlap function $P(q)$ is trivial,
consisting of two delta functions at $\pm q_{\text{EA}}$ in zero
field, whereas in the TNT picture the form of $P(q)$ is quite similar
to the non-trivial (NT) form which Parisi found for the SK model. The
TNT picture accounts for the non-trivial form of the Parisi overlap
function by postulating that there exist droplets of the linear size
$L$ of the system, which contain $O(L^d)$ spins, and which do not
have a free energy of order $L^{\theta}$ (as they would in the droplet
scaling picture), but which have instead a free energy of $O(1)$. It
is the presence of such droplets which makes $P(q)$ non-trivial, which
is a feature observed in all simulations of it to date.

In a recent paper \cite{Moore:21} one of us argued that once the
linear dimension of the system became larger than a crossover length
$L^*$ the non-trivial behavior observed in $P(q)$ will change to the
trivial form predicted by droplet scaling. Estimates of $L^*$ in $d=3$
suggest it might be large, of the order of several hundred lattice
spacings and it is probably the case that to date the regime where
$L> L^*$ has not been reached. Furthermore it was suggested that as
$d \to 6$, $L^*$ would grow towards infinity, as the droplets of
$O(1)$ evolve to the $O(1)$ excitations in the Parisi RSB solution,
where the pure states have free energies which differ from each other
by $O(1)$. In our one dimensional proxy system we would therefore
expect to find that $L^*$ is much larger when $\sigma = 0.75$ than it
is when $\sigma = 0.85$.

In this paper there are TNT-like effects visible in the behavior of
$\xi_{\text{SG}}/N$ in the region where $\xi_{\text{SG}}$ is of $O(N)$
(see Figs. \ref{fig:chi_cl_Nxhr}\flc{e} and
\ref{fig:chi_cl_Nxhr}\flc{f}). When $\xi_{\text{SG}}$ is of order $N$
the droplets which are important are those of size $N$ and if
$L < L^*$ some of these will have free energy of $O(1)$ rather than
$L^{\theta}$. As a consequence the good scaling collapse of the data
visible when $\xi_{\text{SG}}/N \ll 1$ will be lost. In
Figs. \ref{fig:chi_cl_Nxhr_linear}\flc{a} and
\ref{fig:chi_cl_Nxhr_linear}\flc{b} we have plotted
$\xi_{\text{SG}}/N$ on a linear scale versus $N^{1/x}h_r$ focussing
only on the region where $\xi_{\text{SG}}/N$ is of $O(1)$. If the
droplet scaling collapse had been good and of the form of
Eq. (\ref{eq:scalingcorrformxi}) then as $N$ is increased the collapse
should get better and better. In fact due to TNT effects the data in
Fig. \ref{fig:chi_cl_Nxhr_linear}\flc{a} for $\sigma = 0.75$ show the
opposite trend, and the lines get further apart with increasing $N$ in
the region where $\xi_{\text{SG}}$ is of $O(N)$.  However, for
$\sigma = 0.85$ Fig. \ref{fig:chi_cl_Nxhr_linear}\flc{b} shows the
lines seem to be getting closer with increasing $N$. It suggests that
for this value of $\sigma$ we are getting into the region where
$L > L^*$ when droplet scaling applies even when $\xi_{\text{SG}}$ is
of $O(N)$. Data at larger values of $N$ than $4096$ would be nice to
confirm this trend but because these simulations have to be done at
quite low temperatures compared to those for $\sigma = 0.75$ it will
be challenging to do this. Despite this limitation on the size of $N$
which can be reached for $\sigma = 0.85$, there is evidence that for
it, TNT and finite size scaling effects are less troublesome than for
$\sigma = 0.75$, despite the fact that much larger values of $N$ can
be studied at this $\sigma$ value.

\section{Summary and conclusions}
\label{sec:discussion}

In this paper, we have studied the phase transitions in the
one-dimensional power-law diluted XY spin glass, both in the
zero-field limit, and in the presence of a magnetic field random in
the component directions. Whether or not an AT line exists for various
values of the parameter $\sigma$ is a question of fundamental
interest.  To address this, we have performed large scale Monte-Carlo
simulations using a new heatbath algorithm, described in Appendix
\ref{sec:simulation details}. This algorithm hopefully speeds up
equilibration, so cutting computational costs.  We certainly do gain
some advantage in terms of computational time due to the smaller
number of components of XY spins compared to those of the Heisenberg
model. Alas, the heatbath algorithm for XY spins suffers from an
intrinsic disadvantage. Because our algorithm has to generate two
random numbers during each Monte Carlo step, the benefits of the
smaller number of components are largely counterbalanced by the
additional labor involved in the heatbath step. We were unable to go
to larger system sizes than in the corresponding work with Heisenberg
spins \cite{sharma2011almeida}. The largest system sizes that we are
able to simulate are: $N=16384$ for $\sigma=0.6$, $N=32768$ for
$\sigma = 0.75$, while the largest $N$ for $\sigma=0.85$ was
$4096$. The total CPU time spent in generating all the data that we
presented at fixed $h_r$ and varying $T$ was 1183636.2 hrs, which is
135.12 years. The total CPU time consumed in generating the data at
fixed $T$ and varying $h_r$ was 96101.6 days which is 263.29 years.
Thus despite the algorithm not producing significant dividends, we are
able to study fairly large system sizes owing to the expenditure of a
large amount of computer time.

\begin{figure}
  \centering
  \includegraphics[width=0.5\textwidth]{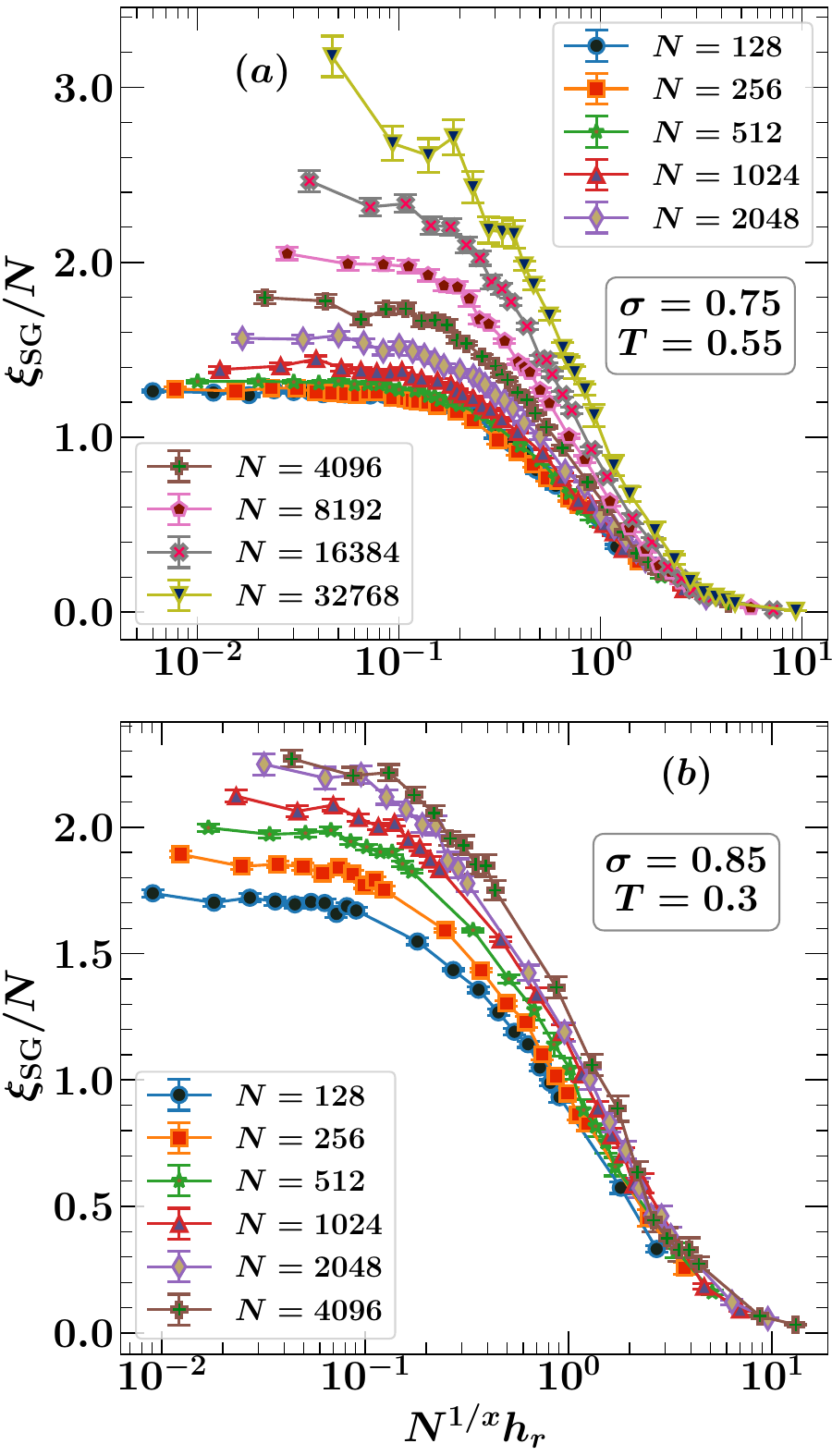}
  \caption{Finite size scaling plots of $\xi_{\text{SG}}$ as a
    function of magnetic field $h_r$, for (a) $\sigma=0.75$ at a
    temperature of $T=0.55$, and (b) $\sigma=0.85$ at a temperature of
    $T=0.3$. These figures show $\xi_{\text{SG}}/N$ plotted on a
    linear scale versus $N^{1/x}h_r$ in the region where
    $\xi_{\text{SG}}/N$ is of $O(1)$}.
  \label{fig:chi_cl_Nxhr_linear}
\end{figure}

The results from our work are broadly in accord with those for the
corresponding Heisenberg spin glass model.  For $\sigma=0.6$, which is
in the mean-field regime, we find a phase transition in the absence of
an external magnetic field, and in the presence of a magnetic field,
which indicates the existence of an AT line. The location of the AT
line is close to the mean-field predictions. For $\sigma=0.75$, which
is in the non-mean-field regime, the conventional data collapse
suggests the existence of an AT line, but the behavior of the
intersections as a function of $N$ indicate that the data is not close
to its large $N$ asymptotic form.  The estimated location of the AT
field based upon intersections that we get from our data at
$\sigma= 0.75$ is strikingly smaller than estimates based on the
mean-field theory formulas. For $\sigma=0.85$, which is deep in the
non-mean-field regime and corresponds to a space dimension of about
$3$, our data are consistent with the absence of an AT line. In this
case there is no crossing of the curves of
$\chi_{\text{SG}}/N^{2-\eta}$ versus $T$ at various $N$ values. But
confusingly intersections $h^*(N, 2N)$, as a function of $h_r$, seem
to exist, whereas intersections $T^*(N, 2N)$ are absent at least for
$\sigma = 0.85$.

However, for $\sigma= 0.75$ and for $\sigma = 0.85$ we found that the
droplet picture provided a much better description of our data from
that obtained assuming the existence of an AT transition line.  The
Imry-Ma formula for the field dependence of $\xi_{\text{SG}}$ works
well until $\xi_{\text{SG}}$ becomes comparable to the system size.  A
similar behavior was reported for the Ising spin glass at
$\sigma = 0.75$ in Ref. \cite{PhysRevE.93.032123}. A finite-size
scaling formulation was developed to treat the data at small fields
when $\xi_{\text{SG}}$ is comparable to the system size $N$, and with
it an excellent collapse of all our data on $\xi_{\text{SG}}$ and
$\chi_{\text{SG}}$ was obtained. We showed that droplet scaling
predicts the existence of the intersections $h^*(N, 2N)$. Our data
unfortunately does not extend to values of $N$ large enough to be in
the asymptotic region where the $N$-dependence of $h^*(N, 2N)$ is
simple. Fortunately there exists a way of testing whether the
intersections are due to an AT transition or are just those predicted
by droplet scaling, which is to study the $N$ dependence of
$R=\chi_{\text{SG}}/N^{2\sigma -1}$, calculated at $h^*(N, 2N)$, and
this test supports the droplet picture provided $N > 1024$ at
$\sigma = 0.75$. Thus it is only for large systems that one can obtain
good evidence for the droplet picture.

We now summarize our main results. The strongest evidence for droplet
scaling is the success of the Imry-Ma formula for the field dependence
of $\xi_{\text{SG}}$ for $\sigma = 0.75$ and $\sigma = 0.85$ (see
inset Figs. \ref{fig:chi_cl_Nxhr}\flc{e} and
\ref{fig:chi_cl_Nxhr}\flc{f}). If droplet scaling works, then no AT
line is to be expected. When $\xi_{\text{SG}} \sim N$ there are
visible sizeable corrections to the Imry-Ma formula which are related
to TNT effects. However for $\sigma = 0.85$ there is tentative
evidence in Fig. \ref{fig:chi_cl_Nxhr_linear}\flc{b} that if even
larger systems could be studied then the TNT effects might be absent,
and so there could exist a length scale $L^*$ above which TNT effects
become unimportant (see Ref. \cite{Moore:21}). If instead of droplet
scaling one assumes that there is an AT phase transition then the
usual finite size scaling plots used to determine $h_{\text{AT}}$ as
in Fig. \ref{fig:chi_cl_vs_hr_s0.75}\flc{c} for $\sigma = 0.75$ are
unsatisfactory: for example the values of $h_{\text{AT}}$ which would
be derived from the crossings of $\xi_{\text{SG}}$ and
$\chi_{\text{SG}}$ as $N$ becomes large look to be significantly
different. In the equivalent data plot for $\sigma = 0.60$ (see
Fig. ~\ref{fig:chi_cl_vs_hr_s0.60}\flc{c}) they are in good
agreement. Furthermore the quantity $R$ of Eq. (\ref{eq:Rdef}) should
approach a constant as $N\to \infty$ if there is a genuine AT
transition, but instead for the cases $\sigma =0.75$
(Fig.~\ref{fig:Rxlog}\flc{b}) and $0.85$
(Fig.~\ref{fig:Rxlog}\flc{c}), it is increasing with $N$ once $N$
becomes large enough.

The simulations of this paper provide numerical evidence that the AT
line and hence RSB is absent in spin glasses below six
dimensions. What is now needed is an explanation of why this might be
the case. Better still would be a rigorous proof that the lower
critical dimension for replica symmetry breaking is six. Our work
indicates that showing that $\sigma = 2/3$ is the precise value of the
critical value of $\sigma $ will be challenging using simulations as
finite size effects are large in its vicinity.

\section*{Acknowledgments}
We are grateful to the High Performance Computing (HPC) facility at
IISER Bhopal, where large-scale calculations in this project were
run. We thank Peter Young and Dan Stein for helpful discussions. B.V
is grateful to the Council of Scientific and Industrial Research
(CSIR), India, for his PhD fellowship. A.S acknowledges financial
support from SERB via the grant (File Number: CRG/2019/003447), and
from DST via the DST-INSPIRE Faculty Award
[DST/INSPIRE/04/2014/002461].

\appendix
\section{The simulation method}
\label{sec:simulation details}

We now give some technical aspects of how the
simulations are run. In the simulations we start with a random initial
configuration and allow it to evolve according to the prescription
given in this section. To incorporate parallel tempering, we
simultaneously simulate $N_T$ copies of the system over $N_T$
different temperatures ranging from $T_{\text{min}}\equiv T_1$ to
$T_{\text{max}}\equiv T_{N_T}$.  In order to facilitate the
computation of the observables outlined in this section, it is
convenient to simulate $4$ sets of $N_T$ copies (2 for $h_r=0$), which
we label (1),(2),(3), and (4).  We perform overrelaxation, heatbath
and parallel tempering sweeps over all these copies keeping track of
the labels appropriately. For every $10$ overrelaxation sweeps we
perform $1$ heatbath and $1$ parallel tempering sweep, since the
overrelaxation sweep involves a significantly lower computational
cost, and is known to speed up equilibration. The parameters of the simulations are shown in Tables \ref{tab:parameters_T} and \ref{tab:parameters_hr}. Once the system
reaches equilibrium, we perform the same number of sweeps in the
measurement phase, so $N_{\text{sweep}}$ is the total number of sweeps
over which the simulation is run, inclusive of both the equilibration
and measurement phases. The last column in the table shows the amount
of computer time expended to generate the data corresponding to the
parameters in that row.  In the measurement phase, we perform one
measurement on the system for every $4$ sweeps. 
The following sections  contain the details of our Monte Carlo simulation procedures.
In order to equilibrate the system as quickly as possible, we perform three
kinds of sweeps: overrelaxation or microcanonical sweeps, heatbath
sweeps, and parallel tempering sweeps.

\begin{figure}
  \includegraphics[width=0.5\textwidth]{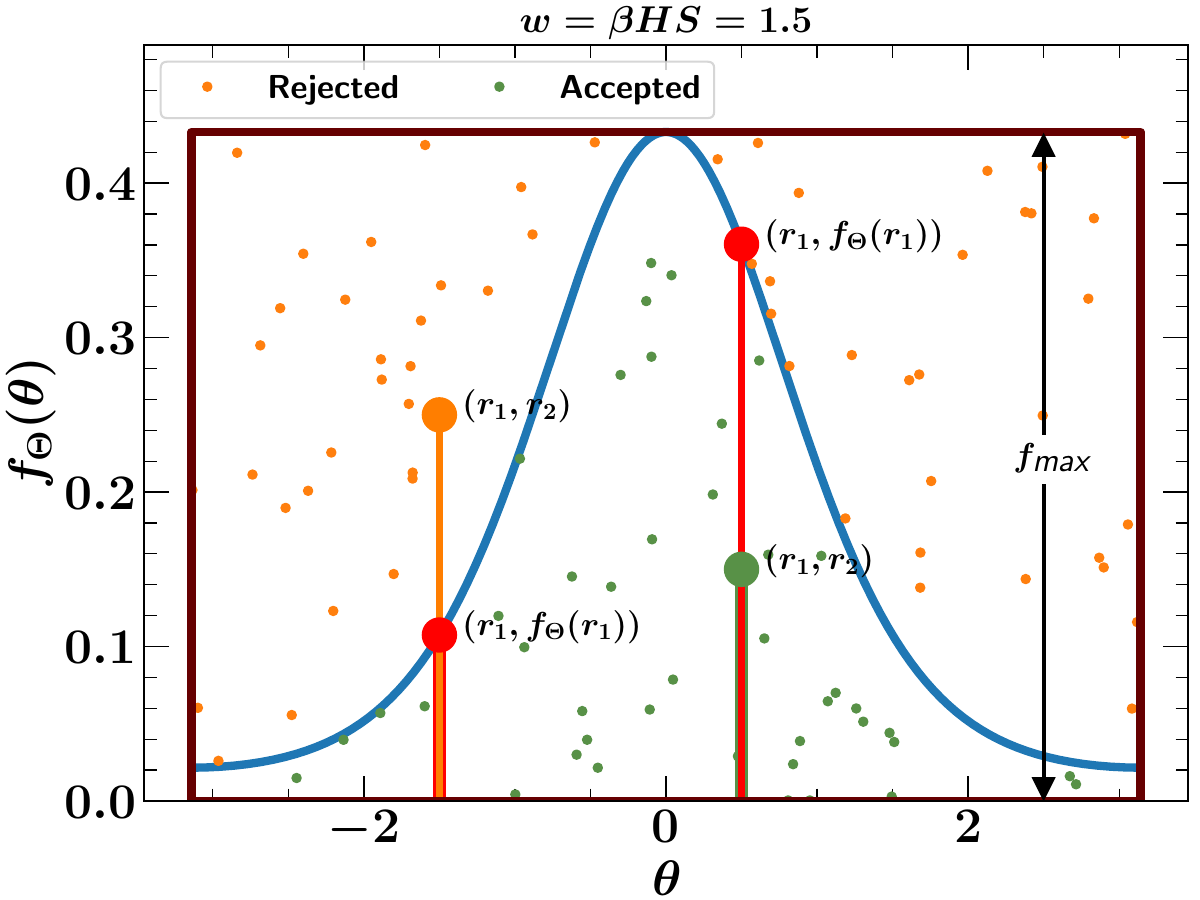}
  \caption{Graphical representation of the rejection method. We
    randomly pick a point $(r_1,r_2)$ within the rectangle from a
    uniform distribution.  If the point lies under the
    $f_{\Theta}(\theta)$ curve given by Eq.~(\ref{eq:xy-pdf}), then
    the point is accepted, and $\theta$ is taken to be $r_1$.
    Otherwise, the point is rejected.}
  \label{fig:rejection_method}
\end{figure}

\subsection{Overrelaxation sweep}
\label{sec:overrelaxation-sweep}

We sweep sequentially through all the lattice sites and compute the
local field $\mathbf{H}_i=\sum_j J_{ij}\mathbf{S}_j + \mathbf{h}_i$ at
a particular lattice site.  The new spin direction $\mathbf{S}'_i$ at
the $i^{\text{th}}$ lattice site is taken to be the mirror image of
the vector $\mathbf{S}_i$ about $\mathbf{H}_i$, i.e.,
\begin{equation}
  \label{eq:or-sweep}
  \mathbf{S}'_i=-\mathbf{S}_i + 2\frac{\mathbf{S}_i\cdot \mathbf{H}_i}{H_i^2}\mathbf{H}_i.
\end{equation}
Since $\mathbf{S}'_i \cdot \mathbf{H}_i = \mathbf{S}_i \cdot
\mathbf{H}_i$, the energy of the system does not change due to these
sweeps.  Hence these sweeps are also called microcanonical sweeps.
These sweeps help us in sampling out the microstates with the same
energy.  The process of equilibration speeds up when we include
overrelaxation sweeps along with the other
sweeps~\cite{PhysRevB.78.014419,PhysRevB.53.2537}.


\subsection{Heatbath sweep}
The overrelaxation sweeps generate states with the same energy and
hence they cannot directly equilibrate the system.  Therefore, we also
perform a heatbath sweep for every $10$ microcanonical sweeps.
Similar to the microcanonical case, we sweep sequentially through the
lattice.

To equilibrate the system, the angle $\theta$ between $\mathbf{H_i}$
and $\mathbf{S^{'}_i}$
should be sampled out from the Boltzmann distribution given by
\begin{equation}
  \label{eq:xy-pdf}
  f_{\Theta}(\theta)=\frac{\text{e}^{-\beta E_i}}{Z}
  =\frac{\text{e}^{\beta H_iS_i\cos\theta}}{Z}=\frac{\text{e}^{w\cos\theta}}{Z} \,,
\end{equation}
where $w=\beta H_i S_i$ and
\begin{equation}
  \label{eq:normalizing_constant}
  Z=\int\limits_{-\pi}^{\pi} \text{e}^{\beta H_iS_i\cos\theta} \,d\theta
\end{equation}
is the normalizing constant. The simplest way to do this is to equate
the cumulative density function (CDF) of $\theta$,
$F_{\Theta}(\theta)$, to that of a uniform distribution:
\begin{equation}
  \label{eq:cdf}
  F_{\Theta}(\theta) = \int\limits_{-\pi}^{\theta} f_{\Theta}(\theta') \,d\theta'
  = \Pi(r_1)=r_1,
\end{equation}
where $r_1$ is a random variable sampled from a uniform distribution
in the interval $(0,1)$.  The value of $\theta$ can be obtained by
simply inverting this function to get
\begin{equation}
  \label{eq:cdf_inverse}
  \theta=F_{\Theta}^{-1}(r_1).
\end{equation}
This method works well with the Heisenberg spins as
$f_{\Theta}(\theta)$ is integrable, which gives an invertible CDF
$F_{\Theta}(\theta)$~\cite{sharma2011almeida,lee2007large}. Since the
probabililty density function (PDF) $f_{\Theta}(\theta)$ for the XY
spin glasses given by Eq.~(\ref{eq:xy-pdf}) is not exactly integrable,
this method cannot be used.

To overcome this problem and to sample out $\theta$ from the Boltzmann
distribution (Eq.~(\ref{eq:xy-pdf})) in as few a number of sweeps as
possible, we develop a heatbath sweep based on the rejection
method~\cite{larson1981}
We generate two random numbers $r_1 \in \text{uniform}(-\pi,\pi)$ and
$r_2 \in \text{uniform}(0,f_{\text{max}})$.  If
$r_2 < f_{\theta}(r_1)$, we accept the move, i.e., take $\theta =r_1$.
Else, we reject the move and generate another pair of random numbers
$(r_1,r_2)$.  This process is repeated until we find an acceptable
value of $r_1$.  A graphical representation for this method is shown
in Fig.~\ref{fig:rejection_method}.  The new spin direction
$\mathbf{S}'_i$ in Cartesian co-ordinates is given by:
\begin{subequations}
  \label{eq:S'_i}
  \begin{align}
    \label{eq:S'_x}
    S'_x &= \cos(\theta+\theta_H), \\
    \label{eq:S'_y}
    S'_y &= \sin(\theta+\theta_H),
  \end{align}
\end{subequations}
where $\theta_H$ is the angle made by the $\mathbf{H}_i$ vector with
the $X$-axis.  Since the generation of random numbers is involved, this
sweep is computationally costlier than others.  Hence we perform more
microcanonical sweeps than heatbath sweeps.

\begin{table*}[t]
   \centering    
   \caption{Parameters of the simulations.  $N_{\text{samp}}$ is the
     number of disorder samples, $N_{\text{sweep}}$ is the number of
     over-relaxation Monte Carlo sweeps for a single disorder sample.
     The system is equilibrated over the first half of the sweeps, and
     measurements are done over the last half of the sweeps with a
     measurement performed every four over-relaxation
     sweeps. $T_{\text{min}}$ and $T_{\text{max}}$ are the lowest and
     highest temperatures simulated, and $N_T$ is the number of
     temperatures used for parallel tempering.}
   \label{tab:parameters_T}
   \begin{ruledtabular}
     \begin{tabular}{llrrrllll}
       $\sigma$  &   $h_r$ &   $N$ &   $N_{\text{samp}}$ &   $N_{\text{sweep}}$ &   $T_{\text{min}}$ &   $T_{\text{max}}$ &   $N_T$ & $t_{\text{tot}}$(hrs) \Tstrut\\[0.15cm]
       \hline
       0.6     &       0 &   128 &        10000 &           512 &        0.6  &        1    &      18 &             0.49 \Tstrut\\
       0.6     &       0 &   256 &         8000 &          1024 &        0.6  &        1    &      22 &             2.23 \\
       0.6     &       0 &   512 &         6400 &          2048 &        0.6  &        1    &      22 &             6.46 \\
       0.6     &       0 &  1024 &         8000 &          4096 &        0.6  &        1    &      26 &            40.74 \\
       0.6     &       0 &  2048 &         3840 &          8192 &        0.6  &        1    &      24 &           105.41 \\
       0.6     &       0 &  4096 &         3200 &         16384 &        0.6  &        1    &      27 &           571.49 \\
       0.6     &       0 &  8192 &         3200 &         32768 &        0.6  &        1    &      30 &          3776.85 \\
       0.6     &       0 & 16384 &         2600 &         65536 &        0.64 &        0.98 &      32 &         18225.5  \\[0.12cm]
       0.6     &     0.1 &   128 &         9600 &          2048 &        0.5  &         0.8 &      21 &             7.75 \\
       0.6     &     0.1 &   256 &         9600 &          2048 &        0.5  &         0.8 &      21 &            15.89 \\
       0.6     &     0.1 &   512 &         9600 &          8192 &        0.5  &         0.8 &      22 &            85.67 \\
       0.6     &     0.1 &  1024 &         8000 &         16384 &        0.5  &         0.8 &      22 &           414.47 \\
       0.6     &     0.1 &  2048 &         7200 &         32768 &        0.5  &         0.8 &      26 &          2029.23 \\
       0.6     &     0.1 &  4096 &         7200 &         65536 &        0.5  &         0.8 &      24 &         10014.8  \\
       0.6     &     0.1 &  8192 &         4380 &        131072 &        0.55 &         0.8 &      25 &         34810.6  \\
       0.6     &     0.1 & 16384 &         7128 &        262144 &        0.55 &         0.8 &      28 &        224425    \\[0.12cm]
       
       0.75    &       0 &   128 &        12800 &          1024 &        0.35 &        0.85 &      21 &             1.6  \\
       0.75    &       0 &   256 &        12800 &          2048 &        0.35 &        0.85 &      24 &             7.21 \\
       0.75    &       0 &   512 &         8000 &          8192 &        0.35 &        0.85 &      24 &            35.22 \\
       0.75    &       0 &  1024 &         8000 &         16384 &        0.35 &        0.85 &      24 &           196.9  \\
       0.75    &       0 &  2048 &         6400 &         32768 &        0.35 &        0.85 &      25 &           774.55 \\
       0.75    &       0 &  4096 &         4880 &         65536 &        0.35 &        0.85 &      27 &          3405.2  \\
       0.75    &       0 &  8192 &         3000 &        131072 &        0.38 &        0.82 &      30 &         14290.9  \\[0.12cm]
       
       0.75    &    0.05 &   128 &        19200 &          8192 &        0.28 &        0.6  &      21 &            45.27 \\
       0.75    &    0.05 &   256 &        16000 &         16384 &        0.28 &        0.6  &      20 &           133.35 \\
       0.75    &    0.05 &   512 &        13600 &         32768 &        0.28 &        0.6  &      20 &           464.77 \\
       0.75    &    0.05 &  1024 &        11000 &         65536 &        0.28 &        0.6  &      21 &          2075.57 \\
       0.75    &    0.05 &  2048 &        10920 &        262144 &        0.28 &        0.6  &      24 &         21314.3  \\
       0.75    &    0.05 &  4096 &        10800 &        524288 &        0.3  &        0.58 &      26 &        123093    \\
       0.75    &    0.05 &  8192 &         5320 &       1048576 &        0.32 &        0.54 &      32 &        364358    \\[0.12cm]
       
       0.85    &       0 &   128 &        12800 &          8192 &        0.2  &        0.5  &      30 &            17.97 \\
       0.85    &       0 &   256 &        12800 &         16384 &        0.2  &        0.5  &      32 &            72.15 \\
       0.85    &       0 &   512 &        12800 &         65536 &        0.2  &        0.5  &      30 &           752.36 \\
       0.85    &       0 &  1024 &        12800 &        131072 &        0.2  &        0.5  &      30 &          3219.93 \\
       0.85    &       0 &  2048 &         8000 &        262144 &        0.2  &        0.5  &      30 &          9504.05 \\
       0.85    &       0 &  4096 &         6480 &        524288 &        0.24 &        0.48 &      30 &         40322.4  \\[0.12cm]
       
       0.85    &    0.02 &   128 &         8000 &         65536 &        0.1  &         0.4 &      30 &           194.33 \\
       0.85    &    0.02 &   256 &         4000 &        131072 &        0.1  &         0.4 &      32 &           470.39 \\
       0.85    &    0.02 &   512 &         4400 &        524288 &        0.1  &         0.4 &      34 &          4780.6  \\
       0.85    &    0.02 &  1024 &         3000 &       2097152 &        0.1  &         0.4 &      35 &         30356.9  \\
       0.85    &    0.02 &  2048 &         1800 &       4194304 &        0.16 &         0.4 &      36 &         84056    \\[0.12cm]
       
       0.85    &    0.05 &   128 &         2000 &         65536 &        0.1  &         0.4 &      30 &            67.07 \\
       0.85    &    0.05 &   256 &         4000 &        131072 &        0.1  &         0.4 &      32 &           604.8  \\
       0.85    &    0.05 &   512 &         3500 &        524288 &        0.1  &         0.4 &      36 &          4958.17 \\
       0.85    &    0.05 &  1024 &         3120 &       2097152 &        0.1  &         0.4 &      36 &         28028.1  \\
       0.85    &    0.05 &  2048 &         3240 &       4194304 &        0.16 &         0.4 &      36 &         151503  \\
     \end{tabular}
   \end{ruledtabular}   
\end{table*}

\begin{table*}[t]
   \centering    
   \caption{Parameters of the simulations done at fixed temperature
     $T$ and varying field $h_r$.  $N(h_r)$ is the number of values of
     field taken in the range $h_r$(min,max). The equilibration times
     are different for different values of the field $h_r$, which lie
     in the range $N_{\text{sweep}}$(min,max). The number of disorder
     samples for different fields lie in the range
     $N_{\text{samp}}$(min,max). $t_{\text{tot}}$ is the total CPU
     time consumed in hours to generate data for a particular system
     size.}
   \label{tab:parameters_hr}
   \begin{ruledtabular}
     \begin{tabular}{llrlrrrl}
       $\sigma$  &   $T$ &   $N$  &   $h_r$(min,max) &  $N(h_r)$  &   $N_{\text{sweep}}$(min,max)  &   $N_{\text{samp}}$(min,max) & $t_{\text{tot}}$(hrs) \Tstrut\\[0.15cm]
       \hline
       0.6     &   0.6 &   128 & $(0.010,9.000)$  &         32 & $(2048,2048)$          & $(2000,64000)$        &            11.44 \Tstrut\\
       0.6     &   0.6 &   256 & $(0.010,9.000)$  &         32 & $(4096,4096)$          & $(2000,48000)$        &            55.5  \\
       0.6     &   0.6 &   512 & $(0.010,9.000)$  &         32 & $(8192,16384)$         & $(2000,80000)$        &           163.91 \\
       0.6     &   0.6 &  1024 & $(0.010,9.000)$  &         32 & $(4096,65536)$         & $(2000,80000)$        &           2375    \\
       0.6     &   0.6 &  2048 & $(0.010,9.000)$  &         32 & $(16384,131072)$       & $(1200,60000)$        &         10259.8  \\
       0.6     &   0.6 &  4096 & $(0.010,9.000)$  &         32 & $(65536,2097152)$      & $(960,28600)$         &         92052.1  \\
       0.6     &   0.6 &  8192 & $(0.010,9.000)$  &         32 & $(65536,4194304)$      & $(400,19428)$         &        310948    \\
       0.6     &   0.6 & 16384 & $(0.010,9.000)$  &         32 & $(131072,4194304)$     & $(488,10689)$         &        1021481    \\[0.12cm]

       0.7     &   0.6 &   128 & $(0.010,9.000)$  &         31 & $(1024,1024)$          & $(4000,4000)$         &             1.86 \\
       0.7     &   0.6 &   256 & $(0.010,9.000)$  &         31 & $(2048,2048)$          & $(1000,8000)$         &             8.53 \\
       0.7     &   0.6 &   512 & $(0.010,9.000)$  &         31 & $(4096,4096)$          & $(1000,8000)$         &            39.46 \\
       0.7     &   0.6 &  1024 & $(0.010,9.000)$  &         31 & $(8192,8192)$          & $(1500,8000)$         &           288.41 \\
       0.7     &   0.6 &  2048 & $(0.010,9.000)$  &         31 & $(16384,16384)$        & $(500,8000)$          &           559.99 \\
       0.7     &   0.6 &  4096 & $(0.010,9.000)$  &         31 & $(32768,32768)$        & $(400,8000)$          &          3020.62 \\
       0.7     &   0.6 &  8192 & $(0.010,9.000)$  &         31 & $(16384,131072)$       & $(2000,6800)$         &         17500.2  \\
       0.7     &   0.6 & 16384 & $(0.010,9.000)$  &         31 & $(32768,262144)$       & $(640,6400)$          &         77734.3  \\[0.12cm]

       0.75    &  0.55 &   128 & $(0.001,9.000)$  &         42 & $(512,1024)$           & $(2000,40000)$        &             6.59 \\
       0.75    &  0.55 &   256 & $(0.001,9.000)$  &         42 & $(1024,2048)$          & $(2000,40000)$        &            56.67 \\
       0.75    &  0.55 &   512 & $(0.001,9.000)$  &         42 & $(4096,4096)$          & $(1000,24000)$        &           165.51 \\
       0.75    &  0.55 &  1024 & $(0.001,9.000)$  &         43 & $(8192,8192)$          & $(1000,12000)$        &           255.75 \\
       0.75    &  0.55 &  2048 & $(0.001,9.000)$  &         43 & $(16384,16384)$        & $(1000,12000)$        &          1000.52 \\
       0.75    &  0.55 &  4096 & $(0.001,9.000)$  &         43 & $(32768,32768)$        & $(800,12000)$         &          5269.11 \\
       0.75    &  0.55 &  8192 & $(0.001,9.000)$  &         42 & $(65536,65536)$        & $(800,8000)$          &         21410.3  \\
       0.75    &  0.55 & 16384 & $(0.001,9.000)$  &         42 & $(131072,131072)$      & $(760,6280)$          &         62062    \\
       0.75    &  0.55 & 32768 & $(0.001,9.000)$  &         42 & $(131072,262144)$      & $(512,4995)$          &        146627    \\[0.12cm]

       0.85    &   0.3 &   128 & $(0.001,9.000)$  &         36 & $(32768,131072)$       & $(2000,30000)$        &           264.97 \\
       0.85    &   0.3 &   256 & $(0.001,9.000)$  &         36 & $(65536,262144)$       & $(1000,25000)$        &          1052.21 \\
       0.85    &   0.3 &   512 & $(0.001,9.000)$  &         36 & $(131072,524288)$      & $(1000,34800)$        &          6161.52 \\
       0.85    &   0.3 &  1024 & $(0.001,9.000)$  &         36 & $(262144,1048576)$     & $(1000,20000)$        &         17668    \\
       0.85    &   0.3 &  2048 & $(0.001,9.000)$  &         36 & $(524288,8388608)$     & $(320,3372)$          &         68933.6  \\
       0.85    &   0.3 &  4096 & $(0.001,9.000)$  &         36 & $(262144,16777216)$    & $(312,5760)$          &        439004    \\
     \end{tabular}
   \end{ruledtabular}   
\end{table*}

\subsection{Parallel tempering sweep}
\label{sec:parall-temp-sweep}

Spin glasses have a complex free energy landscape due to which, at low
temperatures, they tend to get stuck inside metastable valleys, and
true equilibration consumes a lot of time.  At high temperatures, the
system can easily escape the valley due to thermal fluctuations, and
so equilibration is quick. To equilibrate the system in as small a
number of moves as possible, we perform one parallel tempering sweep
for every $10$ overrelaxation
sweeps~\cite{lee2007large,PhysRevB.78.014419}.  To benefit from the
parallel tempering algorithm~\cite{hukushima1996exchange,machta2009strengths}, we simultaneously run the simulation for
$N_T$ copies of the system at $N_T$ different temperatures
$T_1<T_2<T_3<\cdots<T_{N_T}$.  The minimum temperature $T_1$ is the
low temperature at which we are interested in studying the behavior of
the system, and the maximum temperature $T_{N_T}$ is high enough that
the system equilibrates very fast.  We perform overrelaxation and
heatbath sweeps separately on each of the $N_T$ copies of the system.
In the parallel tempering sweep, we compare the energies of two
spin configurations at adjacent temperatures, $T_i$ and $T_{i+1}$,
starting from the smallest temperature $T_1$.  We swap these two spin
configurations such that the detailed balance condition is
satisfied. The Metropolis probability for such a swap is
\begin{align}
  \label{eq:pt-sweep_Metropolis}
  P(T \text{ swap}) &= \min\{1,\exp(\Delta \beta \Delta E)\} \\
  &=
  \begin{cases}
    \exp(\Delta \beta \Delta E),  &(\text{if }\Delta \beta \Delta E < 0),\\
    1, &(\text{otherwise}),
  \end{cases}
\end{align}
where $\Delta \beta=1/T_i-1/T_{i+1}$ and $\Delta E=E_i(T_i)-E_{i+1}(T_{i+1})$.
In this way, a given set of spins performs a random walk in temperature space.

\subsection{Checks for equilibration}
\label{sec:equilibration}

In order to check whether the system has reached equilibrium, we have used a
convenient test~\cite{PhysRevB.63.184422} which is possible because of the
Gaussian nature of the interactions and the onsite external magnetic
field. The relation
\begin{equation}
  \label{eq:equilibration_test}
  U = \frac{zJ^2}{2T}\left(q_l-q_s\right)
  + \frac{h_r^2}{T}\left(q - \left|\textbf{S} \right|^2 \right) ,
\end{equation}
is valid in equilibrium. Here
\begin{align}
  \label{eq:average_energy}
  U &= \frac{1}{N}\left[\langle \mathcal{H} \rangle \right]_{\text{av}} \nonumber \\ 
    &=-\frac{1}{N}\left[\sum\limits_{\langle i,j \rangle} \epsilon_{ij} J_{ij}
        \left\langle \textbf{S}_i \cdot \textbf{S}_j \right\rangle +
        \sum\limits_{i,\mu} h_i^{\mu} \left\langle S_i^{\mu} \right\rangle \right]_{\text{av}} 
\end{align}
is the average energy per spin,
$q=\frac{1}{N} \sum\limits_i \left[ \left\langle \textbf{S}_i \right
  \rangle \cdot \left\langle \textbf{S}_i \right \rangle
\right]_{\text{av}} $ is the Edwards-Anderson order parameter,
$q_l = \frac{1}{N_b}\sum_{\langle i,j \rangle}
\left[\epsilon_{ij}\left\langle \textbf{S}_i\cdot
    \textbf{S}_j\right\rangle^2\right]_{\text{av}} $ is the ``link
overlap'', and
$q_s = \frac{1}{N_b}\sum_{\langle i,j \rangle}
\left[\epsilon_{ij}\left\langle\left( \textbf{S}_i\cdot
      \textbf{S}_j\right)^2 \right\rangle\right]_{\text{av}}$ is the
``spin overlap'', where $N_b=Nz/2$, and $\epsilon_{ij}=1$ if the
$i^{\text{th}}$ and $j^{\text{th}}$ spins are interacting and is zero
otherwise. The $[\cdots]_{\text{av}}$ in Eq.~(\ref{eq:average_energy})
is analytically evaluated by performing integration over $J_{ij}$ and
$h_i^{\mu}$~\cite{Bray_1980} since they have Gaussian distributions.
On evaluating this integral using integration by parts, we get
Eq.~(\ref{eq:equilibration_test}).  As the system reaches equilibrium,
the two sides of Eq.~(\ref{eq:equilibration_test}) approach their
common equilibrium value from opposite directions.

In simulations, we evaluate both sides of
Eq.~(\ref{eq:equilibration_test}) for different number of Monte-Carlo
sweeps (MCSs), which increase in an exponential manner, each value
being twice the previous one.  The averaging is done over the last
half of the sweeps.  We initially start with a random spin
configuration, so the LHS of Eq.~(\ref{eq:equilibration_test}) is
small and the RHS is very large.  As the system gets closer to
equilibrium, these two values come closer to each other from opposite
directions.  When we notice that the averaged quantities satisfy
Eq.~(\ref{eq:equilibration_test}) within error bars, consistently for
at least the last two points, we declare that our system has reached
equilibrium. Once the system reaches equilibrium, we perform the same
number of sweeps in the measurement phase, where we evaluate different
quantities (given below) used to study the possible phase transitions of the
system.


\bibliography{refs}

\end{document}